# A High-Order Relativistic Two-Fluid Electrodynamic Scheme with Consistent Reconstruction of Electromagnetic Fields and a Multidimensional Riemann Solver for Electromagnetism

By

Dinshaw S. Balsara[1], Takanobu Amano[2], Sudip Garain[1] and Jinho Kim[1]

[1]Physics Department, University of Notre Dame; (dbalsara@nd.edu, sgarain@nd.edu, jkim46@nd.edu)

[2]Department of Earth and Planetary Science, University of Tokyo, Tokyo 113-0033, Japan; (amano@eps.s.u-tokyo.ac.jp)

**Abstract**

In various astrophysics settings it is common to have a two-fluid relativistic plasma that interacts with the electromagnetic field. While it is common to ignore the displacement current in the ideal, classical magnetohydrodynamic limit, when the flows become relativistic this approximation is less than absolutely well-justified. In such a situation, it is more natural to consider a positively charged fluid made up of positrons or protons interacting with a negatively charged fluid made up of electrons. The two fluids interact collectively with the full set of Maxwell's equations. As a result, a solution strategy for that coupled system of equations is sought and found here. Our strategy extends to higher orders, providing increasing accuracy.

The primary variables in the Maxwell solver are taken to be the facially-collocated components of the electric and magnetic fields. Consistent with such a collocation, three important innovations are reported here. The first two pertain to the Maxwell solver. In our first innovation, the magnetic field within each zone is reconstructed in a divergence-free fashion while the electric field within each zone is reconstructed in a form that is consistent with Gauss' law. In our second innovation, a multidimensionally upwinded strategy is presented which ensures that the magnetic field can be updated via a discrete interpretation of Faraday's law and the electric field can be updated via a discrete interpretation of the generalized Ampere's law. This multidimensional upwinding is achieved via a multidimensional Riemann solver. The multidimensional Riemann solver automatically provides edge-centered electric field components for the Stokes law-based update of the magnetic field. It also provides edge-centered magnetic field components for the Stokes law-based update of the electric field. The update strategy ensures that the electric field is always consistent with Gauss' law and the magnetic field is always divergence-free. This collocation also ensures that electromagnetic radiation that is propagating in a vacuum has both electric and magnetic fields that are exactly divergence-free.

Coupled relativistic fluid dynamic equations are solved for the positively and negatively charged fluids. The fluids' numerical fluxes also provide a self-consistent current density for the update of the electric field. Our reconstruction strategy ensures that fluid velocities always remain sub-luminal. Our third innovation consists of an efficient design for several popular IMEX schemes so that they provide strong coupling between the finite-volume-based fluid solver and the electromagnetic fields at high order. This innovation makes it possible to efficiently utilize high order IMEX time update methods for stiff source terms in the update of high order finite-volume methods for hyperbolic conservation laws. We also show that this very general innovation should extend seamlessly to Runge-Kutta discontinuous Galerkin methods. The IMEX schemes enable us to use large CFL numbers even in the presence of stiff source terms.

Several accuracy analyses are presented showing that our method meets its design accuracy in the MHD limit as well as in the limit of electromagnetic wave propagation. Several stringent test problems are also presented. We also present a relativistic version of the GEM problem, which shows that our algorithm can successfully adapt to challenging problems in high energy astrophysics.

## I) Introduction

Magnetic fields interacting with non-relativistic and relativistic plasmas dominate many aspects of astrophysics, space physics and plasma fusion research. The present paper focuses on relativistic flows that play an important role in astrophysics. The most popular approximation for that problem involves relativistic Magnetohydrodynamics (MHD). Relativistic MHD (RMHD), quite like classical MHD, use ideal or resistive Ohm's law to replace the electric field with terms that depend on the velocity and the magnetic field. The electric field is no longer an independent variable in RMHD which removes certain fast waves in the system (like electromagnetic radiation) and also results in a smaller equation set. This is a very reasonable approximation in the non-relativistic limit because it removes the speed of light from dominating the timestep in classical MHD codes. Indeed, relativistic flows in plasmas take place with velocities that are comparable to the speed of light. Consequently, a code that solves the full set of Maxwell's equations would take timesteps that are quite comparable to the timesteps in an RMHD code. Even so, very successful schemes have been developed for RMHD (e.g. Komissarov [50], Balsara [19], DelZanna *et al*. [31], Gammie *et al*. [37], Komissarov [51], Ryu *et al*. [67], Mignone & Bodo [60], Honkkila & Janhunen [43], DelZanna *et al*. [32], Mignone, Ugliano and Bodo [61], Anton et al. [1], Kim and Balsara [48], Balsara and Kim [22]).

PIC algorithms that combine relativistic particles with Maxwell's equations have indeed been tried (Spitkovsky and Arons [75], Arons *et al*. [4]) and PIC codes have been very proficient at studying microphysical processes like collisionless shocks and reconnection (Sironi and Spitkovsky [69]). However, relativistic PIC codes are required to resolve the particle gyroradius/gyrofrequency on the mesh, making the study of large-scale macrophysical processes somewhat problematical. A large number of particles are needed within a zone to beat down the accuracy in PIC simulations. Furthermore, PIC simulations are susceptible to numerical noise which only declines as the square root of the number of particles. As a result, PIC codes may be unsuitable for macroscopic simulations.

RMHD also has a genuine limitation because the ratio of the Poynting flux to the particle kinetic energy flux remains fixed in ideal RMHD simulations. There are various astrophysical systems like extragalactic jets, gamma-ray bursts and pulsars where ideal RMHD is an undue limitation. An approximation that allows electromagnetic energy to be smoothly converted into particle kinetic energy might be more suitable for such systems. Fully resistive RMHD requires the solution of a very large set of equations (Watanabe and Yokoyama [81], Komissarov [52], Komissarov *et al*. [53], Dumbser and Zanotti [36]) and may provide one possible way out. Recent interest has, however, converged on relativistic two-fluid electrodynamics (Zenitani, Hesse and Klimas [85], [86], Kojima and Oogi [49], Amano and Kirk [2], Barkov *et al*. [25]). Non-relativistic two-fluid and multi-fluid electrodynamics has been successfully developed (Hakim *et al*. [42], Loverich et al. [57], Wang et al. [80]). It is, therefore, natural to think that relativistic two-fluid electrodynamics should also be developed and that it would have its uses in astrophysics where flows can become relativistic. In relativistic two-fluid electrodynamics a positively charged relativistic fluid and a negatively charged relativistic fluid jointly couple to electric and magnetic fields which are evolved consistently with Maxwell's equations. The positively charged fluid is either made up of positrons or protons and interacts with the electromagnetic field via the Lorentz force. Likewise, the negatively charged fluid is made up of electrons and it too interacts with the electromagnetic field via the Lorentz force. Friction terms can also be introduced between the positive and negative fluids to mimic an Ohms law resistivity. The two fluids jointly produce charge and current densities that contribute to the evolution of the electromagnetic field via the full set of Maxwell's equations. Admittedly, this results in a system of equations that is twice as large as the ideal RMHD system, but it also provides greater fidelity with the physics.

In this paper we focus on developing high accuracy schemes for relativistic two-fluid electrodynamics. Our *first innovation* consists of a novel collocation of electric and magnetic fields which naturally gives rise to a novel reconstruction strategy for the electric fields. This reconstruction is such that it always maintains full consistency with Gauss' law. The reconstruction strategy draws on prior work on the divergence-free reconstruction of magnetic fields in MHD simulations (Balsara [6], [7], [8], Balsara & Dumbser [18], Xu *et al*. [83]). Our *second innovation* consists of realizing that multidimensional Riemann solvers (Balsara [12], [13], [15], [17], Balsara, Dumbser and Abgrall [14], Balsara and Dumbser [16], Balsara *et al*. [21]) that have worked so successfully for numerical MHD can also be extended to numerical electrodynamics. As a result, both Faraday's law and the generalized Ampere's law are both updated in a fashion that is consistent with Stoke's law. The upshot is that in a portion of the computational domain where the charge density is zero, the electric field automatically becomes divergence free. Our *third innovation* consists of showing that an efficient strategy exists for coupling implicit-explicit (IMEX) timestepping schemes for stiff source terms with high order finite-volume discretizations of hyperbolic conservation laws. Our strategy for efficiently using IMEX time update is very general and should extend seamlessly to Runge-Kutta discontinuous Galerkin methods. The IMEX schemes enable us to use large CFL numbers even in the presence of stiff source terms. The present method can successfully bridge the MHD limit and the electrodynamic limit, making it very suitable for simulating pulsar winds or gamma-ray bursts or relativistic jets close to the central engine. By contrast, the resistive RMHD system cannot bridge these two limits.

Section II describes the governing equations. Section III documents the reconstruction of the electric field. Section IV presents multidimensional Riemann solvers and shows that they always have a stabilizing effect for electrodynamic simulations. Section V describes a strategy for maintaining sub-luminal flow velocities. It also presents an efficient strategy for using IMEX schemes for stiff source terms within the context of high order finite-volume discretizations. Section VI documents our numerical method in pointwise fashion. Section VII presents accuracy analysis for numerous idealized tests involving electromagnetic radiation in vacuum; circularly polarized Alfven waves and also circularly polarized electromagnetic radiation propagating through a plasma. It is shown that in all instances the method meets its design accuracy. This gives us confidence that the method accurately straddles the MHD limit and also the electrodynamic limit. Section VIII presents several stringent test problems. Section IX presents conclusions.

**II) Governing Equations for Relativistic Two-Fluid Electrodynamics**

A relativistic charged fluid responds to pressure forces and also the Lorentz force. Consequently, if the particles that make up the fluid have charge "$q$" and mass "$m$" and find themselves in a region of space with electric field **E** and magnetic field **B**, the governing equations for such a fluid become

$$\frac{\partial}{\partial t}(\gamma\rho) + \nabla\cdot(\rho\gamma\mathbf{v}) = 0 \tag{2.1}$$

$$\frac{\partial}{\partial t}\left(\rho h \gamma^2 \mathbf{v}\right) + \nabla \cdot \left(\rho h \gamma^2 \mathbf{vv} + P\mathbf{I}\right) = \frac{q}{m}\gamma\rho\left(\mathbf{E} + \mathbf{v}\times\mathbf{B}\right) \tag{2.2}$$

$$\frac{\partial}{\partial t}\left(\rho h \gamma^2 - P\right) + \nabla \cdot \left(\rho h \gamma^2 \mathbf{v}\right) = \frac{q}{m}\gamma\rho\mathbf{v}\cdot\mathbf{E} \tag{2.3}$$

Here $\rho = nm$ is the proper mass density, with "$n$" being the proper number density; $\mathbf{v}$ is the three-velocity of the fluid; $\gamma = 1/\sqrt{1-\mathbf{v}^2}$ is the Lorentz factor; $h = 1 + \Gamma P/\left(\left(\Gamma-1\right)\rho\right)$ is the specific enthalpy; "$P$" is the gas pressure and $\Gamma$ is the polytropic index. We follow the usual convention in relativity and set the speed of light to be unity for the rest of this paper. Written explicitly in three dimensions, eqns. (2.1) to (2.3) constitute a set of five equations.

In relativistic two-fluid electrodynamics, we consider two oppositely charged species of fluids (e.g. electrons and positrons or electrons and protons). Consequently, we need to evolve two mass densities, two momentum densities and two energy densities. In three dimensions, this constitutes a set of ten evolutionary equations. Let the subscript $s \in \{e, p\}$ denote the specie of the fluid so that we write two sets of equations which look like eqns. (2.1) to (2.3) but with different subscripts. Adding the two sets of equations gives us

$$\frac{\partial}{\partial t}\left(\gamma_e \rho_e + \gamma_p \rho_p\right) + \nabla \cdot \left(\rho_e \gamma_e \mathbf{v}_e + \rho_p \gamma_p \mathbf{v}_p\right) = 0 \tag{2.4}$$

$$\frac{\partial}{\partial t}\left(\rho_e h_e \gamma_e^2 \mathbf{v}_e + \rho_p h_p \gamma_p^2 \mathbf{v}_p\right) + \nabla \cdot \left(\rho_e h_e \gamma_e^2 \mathbf{v}_e \mathbf{v}_e + P_e \mathbf{I} + \rho_p h_p \gamma_p^2 \mathbf{v}_p \mathbf{v}_p + P_p \mathbf{I}\right) = \rho_c \mathbf{E} + \mathbf{j}\times\mathbf{B} \tag{2.5}$$

$$\frac{\partial}{\partial t}\left(\rho_e h_e \gamma_e^2 - P_e + \rho_p h_p \gamma_p^2 - P_p\right) + \nabla \cdot \left(\rho_e h_e \gamma_e^2 \mathbf{v}_e + \rho_p h_p \gamma_p^2 \mathbf{v}_p\right) = \mathbf{j}\cdot\mathbf{E} \tag{2.6}$$

On the right hand sides of these equations we can see the charge density and current density vector which are defined very naturally by the densities of the two fluids and their mass fluxes as

$$\rho_c = \frac{q_e}{m_e}\gamma_e \rho_e + \frac{q_p}{m_p}\gamma_p \rho_p \tag{2.7}$$

$$\mathbf{j} = \frac{q_e}{m_e}\gamma_e \rho_e \mathbf{v}_e + \frac{q_p}{m_p}\gamma_p \rho_p \mathbf{v}_p \tag{2.8}$$

It is easy to see that if the fluid variables are given a finite volume interpretation and collocated at the zone-centers then the charge density can be written as a weighted difference between the two oppositely charged fluid densities. Likewise, the mass fluxes from the two continuity equations would provide the appropriate components of the vector of current density that are collocated at the corresponding faces. The charge density and the vector of current density, therefore, provide the source terms for Maxwell's equations for the plasma.

After asserting the constitutive relations and setting the speed of light to unity, Maxwell's equations can be written in the form

$$\frac{\partial}{\partial t}\mathbf{B} = -\nabla \times \mathbf{E} \qquad (2.9)$$

$$\frac{\partial}{\partial t}\mathbf{E} = \nabla \times \mathbf{B} - 4\pi \mathbf{j} \qquad (2.10)$$

$$\nabla \cdot \mathbf{E} = 4\pi \rho_c \qquad (2.11)$$

$$\nabla \cdot \mathbf{B} = 0 \qquad (2.12)$$

The first two equations from (2.9) and (2.10) are commonly known as Faraday's law and the generalized Ampere's law; and both those equations are evolutionary. Consequently, the next two equations – Gauss' law (2.11) and the divergence-free magnetic field (2.12) – can be viewed as constraints. Once the constraints hold on the computational mesh, they should hold for the entire duration of any computation. Notice that eqns. (2.9) and (2.10) add six equations to the evolutionary set of equations that we need to evolve in three dimensions.

The present equation set is also extensible to the case where multiple species of charged particles, for example multiple species of ions, are present. In that case, each ionic specie would have its own mass, momentum and energy equations. Furthermore, each specie would contribute to the charge and current densities in eqns. (2.7) and (2.8). Our present formulation ignores kinetic effects associated with finite Larmor radius and wave-particle resonances. However, effects associated with the finite inertia of each specie is retained.

Faraday's law and the divergence-free constraint for the magnetic field are naturally satisfied on a Yee-type mesh (Yee [84]). A considerable amount of work has been done on Faraday's law within the context of MHD and recent research has focused on two major advances:- First, we have understood the nature of divergence-free magnetic fields and their reconstruction on a mesh (Balsara [6], [7], [8], Balsara & Dumbser [18], Xu *et al*. [83]). Second, we have shown the importance of multidimensional Riemann solvers for obtaining the corresponding multidimensionally upwinded electric field (Balsara [12], [13], [15], [17], Balsara, Dumbser and Abgrall [14], Balsara and Dumbser [16], [18]).

In this paper we demonstrate that it is beneficial to collocate the components of the electric field at the zone faces and treat them as the primary variables of the computation; please see Fig. 1. As a result, the primary electric field variables in the generalized Ampere's law can also be collocated at zone faces. This is the same face-centered collocation that was adopted for the current density, which yields even further benefits in the update of the electric field. By Gauss' law, the reconstructed electric fields within a zone would then have to be consistent with Gauss' law and also match the primary electric field components that are collocated at zone faces. This requires the invention of a novel type of reconstruction strategy for the electric field and such a strategy is presented here. Such a reconstruction strategy is fully documented in Section III of this paper.

Fig. 1 shows us that the primary magnetic and electric field variables are facially-collocated and undergo an update from Faraday's law and the generalized Ampere's law respectively. The components of the primary magnetic field are shown by the thick black arrows while the components of the primary electric field are shown by the thick red arrows. The update consists of asserting Stokes law for the evolutionary equations in each of the faces of the mesh. This necessarily requires obtaining edge-centered electric and magnetic field components that are consistent with Maxwell's equations. The edge-centered electric fields, which are used for updating the facial magnetic field components, are shown by the thin black arrows close to the appropriate edge. The edge-centered magnetic fields, which are used for updating the facial electric field components, are shown by the thin red arrows close to the appropriate edge.

The edge-centered electric and magnetic fields in Fig. 1 are crucial to our update strategy. They have to be obtained with special attention that is devoted to multidimensional upwinding.

As an example, focus for a moment on the edge-centered z-components of the electric or magnetic fields. They can be influenced by electromagnetic radiation that propagates in the x-direction or in the y-direction or, indeed, any direction in the xy-plane. We, therefore, realize that the edge-centered electric and magnetic fields should also be multidimensionally upwinded. Such a task is accomplished in this paper by the invention of a novel, multidimensional Riemann solver for Maxwell's equations. Such Riemann solvers go under the name of MuSIC Riemann solvers; where MuSIC stands for "Multidimensional, Self-similar, strongly-Interacting, Consistent". Such Riemann solvers are multidimensional; they draw on the self-similarity of the problem; they focus on the strongly-interacting state that results when multiple one-dimensional Riemann solvers interact; and the design relies on establishing consistency with the conservation law. For more information on MuSIC Riemann solvers and their application to a range of hyperbolic systems, please see Balsara [12], [13], [15], [17], Balsara, Dumbser and Abgrall [14], Balsara and Dumbser [16]. MuSIC Riemann solvers for Maxwell's equations, as well as their desirable stabilizing properties, are fully documented in Section IV of this paper. Focusing back on Fig. 1 we see that the edge-centered electric and magnetic fields that are shown in that figure have a "*" superscript and all such variables are indeed obtained from the MuSIC Riemann solver.

Eqns. (2.5) and (2.6) will, in general, have very stiff source terms. It is a standard approach (favored even for the MHD equations) to convert the source terms into flux form. This can be done using Maxwell's equations with the result that eqns. (2.5) and (2.6) become

$$\frac{\partial}{\partial t}\left(\rho_e h_e \gamma_e^2 \mathbf{v}_e + \rho_p h_p \gamma_p^2 \mathbf{v}_p + \frac{1}{4\pi}\mathbf{E}\times\mathbf{B}\right)$$
$$+\nabla\cdot\left(\rho_e h_e \gamma_e^2 \mathbf{v}_e\mathbf{v}_e + P_e\mathbf{I} + \rho_p h_p \gamma_p^2 \mathbf{v}_p\mathbf{v}_p + P_p\mathbf{I} + \frac{1}{8\pi}\left(\mathbf{E}^2+\mathbf{B}^2\right)\mathbf{I} - \frac{1}{4\pi}\left(\mathbf{E}\otimes\mathbf{E}+\mathbf{B}\otimes\mathbf{B}\right)\right) = 0 \quad (2.13)$$

and

$$\frac{\partial}{\partial t}\left(\rho_e h_e \gamma_e^2 - P_e + \rho_p h_p \gamma_p^2 - P_p + \frac{1}{8\pi}\left(\mathbf{E}^2+\mathbf{B}^2\right)\right) + \nabla\cdot\left(\rho_e h_e \gamma_e^2 \mathbf{v}_e + \rho_p h_p \gamma_p^2 \mathbf{v}_p + \frac{1}{4\pi}\mathbf{E}\times\mathbf{B}\right) = 0 \quad (2.14)$$

We see that the Poynting flux contributes to the total momentum density as well as to the flux of the total energy density. We also see that the electromagnetic energy density contributes to the total energy density. Furthermore, we can see that the momentum flux is influenced by the electromagnetic pressure and the electromagnetic tension terms. Eqns. (2.13) and (2.14) are our new equations for the evolution of the total momentum density and the total energy density. Since they are in a nice conservation form, they are well-suited for computer implementation. Eqn. (2.4) is still our equation for the evolution of the total mass density.

While eqns. (2.4), (2.13) and (2.14) provide evolutionary equations for the total mass density, the total momentum density and the total energy density, we also need an evolutionary equation for the charge density that is on the right hand side of eqn. (2.11). From eqn. (2.7) please realize that it is just weighted average of the electron and positron/proton charge densities. Therefore, multiplying the continuity equation for the electrons with $q_e/m_e$ and the continuity equation for the oppositely charged positrons/protons with $q_p/m_p$ and adding the resulting equations we get

$$\frac{\partial}{\partial t}\left(\frac{q_e}{m_e}\gamma_e\rho_e + \frac{q_p}{m_p}\gamma_p\rho_p\right) + \nabla\cdot\left(\frac{q_e}{m_e}\rho_e\gamma_e\mathbf{v}_e + \frac{q_p}{m_p}\rho_p\gamma_p\mathbf{v}_p\right) = 0 \qquad (2.15)$$

In most normal circumstances we expect $q_e = -q_p$ so that eqn. (2.15) is a weighted difference between the mass density equations for the negatively and positively charged fluids. Happily, from eqns. (2.7) and (2.8) we see that this is just the familiar equation for the charge density and the current density that is usually written as

$$\frac{\partial}{\partial t}\rho_c + \nabla\cdot\mathbf{j} = 0 \qquad (2.16)$$

Eqns. (2.13) and (2.14) give us the evolution of the sum of the momentum and energy densities. To form a nice set of equations, and also to remain consistent with the underlying philosophy that gave rise to eqn. (2.15), we need equations that express the difference in the momentum and energy densities. This is accomplished by using the same weighting by the charge to mass ratios. We therefore get

$$\frac{\partial}{\partial t}\left(\frac{q_e}{m_e}\rho_e h_e \gamma_e^2 \mathbf{v}_e + \frac{q_p}{m_p}\rho_p h_p \gamma_p^2 \mathbf{v}_p\right)$$
$$+ \nabla \cdot \left(\frac{q_e}{m_e}\rho_e h_e \gamma_e^2 \mathbf{v}_e \mathbf{v}_e + \frac{q_e}{m_e}P_e \mathbf{I} + \frac{q_p}{m_p}\rho_p h_p \gamma_p^2 \mathbf{v}_p \mathbf{v}_p + \frac{q_p}{m_p}P_p \mathbf{I}\right) = \Lambda \mathbf{E} + \mathbf{\Phi} \times \mathbf{B} \quad (2.17)$$

and

$$\frac{\partial}{\partial t}\left(\frac{q_e}{m_e}\rho_e h_e \gamma_e^2 - \frac{q_e}{m_e}P_e + \frac{q_p}{m_p}\rho_p h_p \gamma_p^2 - \frac{q_p}{m_p}P_p\right) + \nabla \cdot \left(\frac{q_e}{m_e}\rho_e h_e \gamma_e^2 \mathbf{v}_e + \frac{q_p}{m_p}\rho_p h_p \gamma_p^2 \mathbf{v}_p\right) = \mathbf{\Phi} \cdot \mathbf{E} \quad (2.18)$$

where we define

$$\Lambda = \left(\frac{q_e}{m_e}\right)^2 \gamma_e \rho_e + \left(\frac{q_p}{m_p}\right)^2 \gamma_p \rho_p \quad (2.19)$$

and

$$\mathbf{\Phi} = \left(\frac{q_e}{m_e}\right)^2 \gamma_e \rho_e \mathbf{v}_e + \left(\frac{q_p}{m_p}\right)^2 \gamma_p \rho_p \mathbf{v}_p \quad (2.20)$$

This completes our description of the relativistic two-fluid electrodynamic system.

Finally, we identify the evolutionary equations that should be implemented in a computer code. In three dimensions we see that eqns. (2.4), (2.13) and (2.14) along with eqns. (2.15), (2.17) and (2.18) give us ten equations in conservation form for evolving the mass, momentum and energy densities for each of the two fluids. In addition to that, eqns. (2.9) and (2.10) give us six more equations that require a Stokes law-based update in the faces of the mesh with a reconstruction of electromagnetic fields that is consistent with eqns. (2.11) and (2.12).

The above system can be extended to include resistive effects. Such terms usually model the friction that arises when a plasma with one charge streams through a plasma with an opposite charge (Zenitani, Hesse and Klimas [85], [86]). The inclusion of such terms, while it can be motivated from a two-stream instability, is mostly phenomenological. The upshot is that eqns. (2.4), (2.13), (2.14) and eqn. (2.15) remain unchanged. However, eqn. (2.17) becomes

$$\frac{\partial}{\partial t}\left(\frac{q_e}{m_e}\rho_e h_e \gamma_e^2 \mathbf{v}_e + \frac{q_p}{m_p}\rho_p h_p \gamma_p^2 \mathbf{v}_p\right)$$
$$+ \nabla \cdot \left(\frac{q_e}{m_e}\rho_e h_e \gamma_e^2 \mathbf{v}_e \mathbf{v}_e + \frac{q_e}{m_e}P_e \mathbf{I} + \frac{q_p}{m_p}\rho_p h_p \gamma_p^2 \mathbf{v}_p \mathbf{v}_p + \frac{q_p}{m_p}P_p \mathbf{I}\right) = \Lambda \mathbf{E} + \mathbf{\Phi} \times \mathbf{B} + \mathbf{R}$$

(2.21)

and

$$\frac{\partial}{\partial t}\left(\frac{q_e}{m_e}\rho_e h_e \gamma_e^2 - \frac{q_e}{m_e}P_e + \frac{q_p}{m_p}\rho_p h_p \gamma_p^2 - \frac{q_p}{m_p}P_p\right) + \nabla \cdot \left(\frac{q_e}{m_e}\rho_e h_e \gamma_e^2 \mathbf{v}_e + \frac{q_p}{m_p}\rho_p h_p \gamma_p^2 \mathbf{v}_p\right) = \mathbf{\Phi} \cdot \mathbf{E} + R^0$$

(2.22)

Here we define $\mathbf{R}$ and $R^0$ as

$$\mathbf{R} = -\eta\left(\frac{\Omega^2}{4\pi}\mathbf{j} - \rho_0 \mathbf{\Phi}\right)$$

where $\Omega^2 \equiv 4\pi\left[\left(\frac{q_p}{m_p}\right)^2 \rho_p + \left(\frac{q_e}{m_e}\right)^2 \rho_e\right]$ and $\rho_0 \equiv \frac{4\pi}{\Omega^2}(\Lambda \rho_c - \mathbf{j} \cdot \mathbf{\Phi})$

(2.23)

and

$$R^0 = -\eta\left(\frac{\Omega^2}{4\pi}\rho_c - \rho_0 \Lambda\right)$$

(2.24)

where $\Omega$ is the plasma frequency and $\rho_0$ is the charge density as measured in the rest frame of the fluid. In the above two equations $\eta$ is a parameter that parametrizes the resistivity. With the above definitions, it has been shown by Amano [3] that one can retrieve the resistive MHD limit of Komissarov [52].

### III) Reconstruction of the Electric Field

When the charge density is zero, Gauss' law tells us that the electric field is divergence-free. As a result, the reconstruction of the electric field is somewhat like the divergence-free reconstruction for magnetic fields that was invented in Balsara [6], [7], [8], Balsara & Dumbser [18] and Xu *et al*. [83]. However, from an application of Gauss' law we notice that the divergence of the electric field is indeed proportional to the charge density. The electric charge is a zone-averaged quantity; as a result, its higher moments can be obtained via a standard finite volume reconstruction process. Consequently, the higher moments of the divergence of the electric field must match the higher moments of the charge. We illustrate this concept at second, third and fourth orders in this paper. Once the concept is grasped, it can be extended to all other orders and on any mesh type (Balsara & Dumbser [18]). In the rest of this section we describe the second order accurate case. The second order reconstruction is easy to understand and it also helps the reader by cataloguing the notation. We then describe the third order accurate case in Appendix A and the fourth order case in Appendix B.

Let the mean electric field's x-components in the right and left x-faces of a unit cube (reference element) with extent $[-1/2, 1/2]^3$ be denoted by $E_0^{x\pm}$ respectively. Similarly, let the mean electric field's y-components in the upper and lower y-faces of the unit cube be denoted by $E_0^{y\pm}$ respectively. Furthermore, let the mean electric field's z-components in the top and bottom z-faces of the unit cube be denoted by $E_0^{z\pm}$ respectively. The x-component of the electric field in the right and left x-faces also has moments $E_y^{x\pm}$ and $E_z^{x\pm}$ in the y- and z-directions. Similarly, the y-component of the electric field in the upper and lower y-faces also has moments $E_x^{y\pm}$ and $E_z^{y\pm}$ in the x- and z-directions. Likewise, the z-component of the electric field in the top and bottom z-faces also has moments $E_x^{z\pm}$ and $E_y^{z\pm}$ in the x- and y-directions. Therefore, at the right and left x-faces of the reference element, the reconstruction in the interior will have to match the two linear profiles for the x-component of the electric field given by

$$E^{x\pm}(y,z) = E_0^{x\pm} + E_y^{x\pm} y + E_z^{x\pm} z \qquad (3.1)$$

The linear profiles in the above equation can be obtained by using TVD or WENO reconstruction within the x-face of interest by using the x-components of the neighboring electric fields in the neighboring x-faces. The two linear profiles for the y-component of the electric field at the upper and lower y-faces are given by

$$E^{y\pm}(x,z) = E_0^{y\pm} + E_x^{y\pm} x + E_z^{y\pm} z \qquad (3.2)$$

Similarly, the two linear profiles for the z-component of the electric field at the top and bottom z-faces are given by

$$E^{z\pm}(x,y) = E_0^{z\pm} + E_x^{z\pm} x + E_y^{z\pm} y \qquad (3.3)$$

All these profiles have to be matched in the faces by the reconstructed electric field within the reference element.

Since the present sub-section focuses on the second order case, the charge density is reconstructed in piecewise linear fashion as

$$\rho_c(x,y,z) = q_0 + q_x x + q_y y + q_z z \qquad (3.4)$$

Let the electric field within the unit cube be described by the following polynomials

$$E^x(x,y,z) = a_0 + a_x x + a_y y + a_z z + a_{xx}(x^2 - 1/12) + a_{xy} x y + a_{xz} x z \qquad (3.5a)$$

$$E^y(x,y,z) = b_0 + b_x x + b_y y + b_z z + b_{xy} x y + b_{yy}(y^2 - 1/12) + b_{yz} y z \qquad (3.5b)$$

$$E^z(x,y,z) = c_0 + c_x x + c_y y + c_z z + c_{xz} x z + c_{yz} y z + c_{zz}(z^2 - 1/12) \qquad (3.5c)$$

The logic for the selection of these polynomials has been explained in Balsara [6]. For the sake of simplicity, we assume that factors of $4\pi$ etc. in Gauss' law have been absorbed in the definition of the charge density. As a result, we can write Gauss' law in a very compact form as

$$\partial_x E^x(x,y,z) + \partial_y E^y(x,y,z) + \partial_z E^z(x,y,z) = \rho_c(x,y,z) \qquad (3.6)$$

Please compare eqn. (3.6) to eqn. (2.11) to see how the factor of $4\pi$ has been absorbed in the charge density just because of notational convenience; i.e. the transcriptions $\rho \to 4\pi \rho$ and $q \to 4\pi q$ would restore the appropriate factor of $4\pi$ in all the equations in this Section. By considering the linear variations in Gauss' law we get the following consistency relations

$$2a_{xx} + b_{xy} + c_{xz} = q_x \quad ; \quad a_{xy} + 2b_{yy} + c_{yz} = q_y \quad ; \quad a_{xz} + b_{yz} + 2c_{zz} = q_z \qquad (3.7)$$

By considering the constant terms in Gauss' law we get the following consistency relation

$$a_x + b_y + c_z = q_0 \qquad (3.8)$$

In the next paragraph we will find the coefficients that satisfy these consistency relations.

Matching the linearly varying part of the x-component of the electric field in the right and left x-faces gives

$$a_y = \left(E_y^{x+} + E_y^{x-}\right)/2 \quad ; \quad a_{xy} = E_y^{x+} - E_y^{x-} \quad ; \quad a_z = \left(E_z^{x+} + E_z^{x-}\right)/2 \quad ; \quad a_{xz} = E_z^{x+} - E_z^{x-}$$

$$(3.9)$$

Matching the linearly varying part of the y-component of the electric field in the upper and lower y-faces gives

$$b_x = \left(E_x^{y+} + E_x^{y-}\right)/2 \quad ; \quad b_{xy} = E_x^{y+} - E_x^{y-} \quad ; \quad b_z = \left(E_z^{y+} + E_z^{y-}\right)/2 \quad ; \quad b_{yz} = E_z^{y+} - E_z^{y-}$$

$$(3.10)$$

Matching the linearly varying part of the z-component of the electric field in the top and bottom z-faces gives

$$c_x = \left(E_x^{z+} + E_x^{z-}\right)/2 \quad ; \quad c_{xz} = E_x^{z+} - E_x^{z-} \quad ; \quad c_y = \left(E_y^{z+} + E_y^{z-}\right)/2 \quad ; \quad c_{yz} = E_y^{z+} - E_y^{z-}$$

$$(3.11)$$

The consistency relations imposed by Gauss' law now enable us to write

$$a_{xx} = -\left(b_{xy} + c_{xz} - q_x\right)/2 \quad ; \quad b_{yy} = -\left(a_{xy} + c_{yz} - q_y\right)/2 \quad ; \quad c_{zz} = -\left(a_{xz} + b_{yz} - q_z\right)/2 \qquad (3.12)$$

Matching the constant part of the x-component of the electric field in the right and left x-faces gives

$$a_0 = \left(E_0^{x+} + E_0^{x-}\right)/2 - a_{xx}/6 \quad ; \quad a_x = E_0^{x+} - E_0^{x-} \qquad (3.13)$$

Matching the constant part of the y-component of the electric field in the upper and lower y-faces gives

$$b_0 = \left(E_0^{y+} + E_0^{y-}\right)/2 - b_{yy}/6 \quad ; \quad b_y = E_0^{y+} - E_0^{y-} \tag{3.14}$$

Matching the constant part of the z-component of the electric field in the top and bottom z-faces gives

$$c_0 = \left(E_0^{z+} + E_0^{z-}\right)/2 - c_{zz}/6 \quad ; \quad c_z = E_0^{z+} - E_0^{z-} \tag{3.15}$$

It is also easy to see that

$$a_x + b_y + c_z = \left(E_0^{x+} - E_0^{x-}\right) + \left(E_0^{y+} - E_0^{y-}\right) + \left(E_0^{z+} - E_0^{z-}\right) = q_0 \tag{3.16}$$

I.e., the above equation is an expression of Gauss' law for the lowest order terms. This completes our description of the second order accurate reconstruction of the electric field in the presence of a charge density. When the charge density and its variations are exactly zero, we retrieve the divergence-free reconstruction that is so useful for specifying the magnetic field within a zone. When making a computer implementation, it is best to implement the equations in the sequence that they have been described here.

**IV) A Multidimensional Riemann Solver for Electrodynamics**

This section is divided into two parts. The first sub-section catalogues the electric and magnetic field components that arise via the application of a multidimensional Riemann solver. The second sub-section shows explicitly that the multidimensional Riemann solver always has a stabilizing influence in the limit of first order of accuracy. This gives us some confidence that the resulting method is always stable.

**IV.a) Cataloguing the Electric and Magnetic fields from the Multidimensional Riemann Solver**

The evolutionary part of the electrodynamic system, i.e. eqns. (2.9) and (2.10), can be written in conservation form as

$$\frac{\partial}{\partial t}\begin{pmatrix} E^x \\ E^y \\ E^z \\ B^x \\ B^y \\ B^z \end{pmatrix} + \frac{\partial}{\partial x}\begin{pmatrix} 0 \\ cB^z \\ -cB^y \\ 0 \\ -cE^z \\ cE^y \end{pmatrix} + \frac{\partial}{\partial y}\begin{pmatrix} -cB^z \\ 0 \\ cB^x \\ cE^z \\ 0 \\ -cE^x \end{pmatrix} + \frac{\partial}{\partial z}\begin{pmatrix} cB^y \\ -cB^x \\ 0 \\ -cE^y \\ cE^x \\ 0 \end{pmatrix} = \begin{pmatrix} -4\pi j_x \\ -4\pi j_y \\ -4\pi j_z \\ 0 \\ 0 \\ 0 \end{pmatrix}$$ (4.1)

The speed of light "$c$" is retained as-such in this Sub-section; but one could just as well follow the relativistic convention and set it to unity. Only the homogeneous part is relevant to the construction of a Riemann solver, which is we why we focus on it here. Consequently, the right hand side of eqn. (4.1) can be set to zero for the purposes of the analysis presented here. The extremal speeds in any one direction are set by the speed of light, i.e., "$-c$" and "$c$", with the result that any approximate one-dimensional Riemann solver would be a Lax-Freidrichs Riemann solver. Using eqns. (12), (13) and (14) of Balsara [15], we can obtain the solution of the multidimensional Riemann problem. The PDEs for the update of the electric and magnetic fields are based on Stokes law, and the utility of the multidimensional Riemann solver in such situations has been amply documented in the literature (Balsara [12], [13], [15], [17], Balsara, Dumbser and Abgrall [14], Balsara and Dumbser [16], [18]).

Fig. 2 shows four zones in the xy-plane that come together at the z-edge of a three-dimensional mesh. Since the mesh is viewed from the top in plan view, the z-edge is shown by the black dot and the four abutting zones are shown as four squares. In other words, we are looking down the z-axis which is taken to coincide with the z-edge of the mesh. The divergence-free reconstruction is used in each of the four zones shown, with the result that there are four states that come together at the edge. The four states have subscripts given by "RU" for right-upper; "LU" for left-upper; "LD" for left-down and "RD" for right-down. Electric and magnetic fields are reconstructed in the four zones shown in Fig. 2 using the reconstruction strategy described in Section III. The reconstructed electric and magnetic fields are then evaluated at the location of the z-edge in Fig. 2 where we desire the electric field. These four states constitute the four input states for the multidimensional Riemann problem. The multidimensional Riemann problem then produces the edge-centered z-components of the electric and magnetic fields. These are the very same fields with a "*" for a superscript that are shown in Fig. 1. Please also realize that while the x-components of the electric and magnetic fields will be continuous across an x=constant boundary of the mesh; the y-components of the electric and magnetic fields can indeed have a jump across that boundary. Similarly, while the y-components of the electric and magnetic fields will be continuous across a y=constant boundary of the mesh; the x-components of the electric and magnetic fields can indeed have a jump across that boundary. It is via this feature of the reconstruction that we ensure that appropriate amounts of truly multidimensional dissipation are provided to stabilize the scheme.

The z-component of the electric field that is provided by the multidimensional Riemann solver can then be written as

$$E^{z*} = \frac{1}{4}\left(E^z_{RU} + E^z_{RD} + E^z_{LU} + E^z_{LD}\right) + \frac{1}{2}\left(B^{y*}_R - B^{y*}_L\right) - \frac{1}{2}\left(B^{x*}_U - B^{x*}_D\right) \tag{4.2}$$

In order to obtain the most symmetrical expressions, we can write $B^{y*}_R = \left(B^y_{RU} + B^y_{RD}\right)/2$, $B^{y*}_L = \left(B^y_{LU} + B^y_{LD}\right)/2$, $B^{x*}_U = \left(B^x_{RU} + B^x_{LU}\right)/2$ and $B^{x*}_D = \left(B^x_{RD} + B^x_{LD}\right)/2$. Strictly speaking, if the consistent reconstruction is used we should have $B^y_{RU} = B^y_{RD}$ because it is just an assertion that the y-component of the magnetic field is single-valued in the y-face on the right side of Fig. 2; we should also have $B^y_{LU} = B^y_{LD}$ because it is just an assertion that the y-component of the magnetic field is single-valued in the y-face on the left side of Fig. 2. Similar considerations applied to the x-faces of Fig. 2 give $B^x_{RU} = B^x_{LU}$ and $B^x_{RD} = B^x_{LD}$.

The z-component of the magnetic field that is provided by the multidimensional Riemann solver can then be written as

$$B^{z*} = \frac{1}{4}\left(B^z_{RU} + B^z_{RD} + B^z_{LU} + B^z_{LD}\right) - \frac{1}{2}\left(E^{y*}_R - E^{y*}_L\right) + \frac{1}{2}\left(E^{x*}_U - E^{x*}_D\right) \tag{4.3}$$

In order to obtain the most symmetrical expressions, we can again write $E^{y*}_R = \left(E^y_{RU} + E^y_{RD}\right)/2$, $E^{y*}_L = \left(E^y_{LU} + E^y_{LD}\right)/2$, $E^{x*}_U = \left(E^x_{RU} + E^x_{LU}\right)/2$ and $E^{x*}_D = \left(E^x_{RD} + E^x_{LD}\right)/2$. As before, from Fig. 2 a consistent reconstruction would indeed give us $E_{RU}^y = E_{RD}^y$ , $E_{LU}^y = E_{LD}^y$ , $E_{RU}^x = E_{LU}^x$ and $E_{RD}^x = E_{LD}^x$ .

Cyclic rotations can be applied to eqns. (4.2) and (4.3) to obtain the x- and y-components of the electric and magnetic fields from the multidimensional Riemann solver. Notice that eqns. (4.2) and (4.3) catalogue the z-components of the electric and magnetic fields as a centered part plus a dissipative contribution. Also realize that as the order of accuracy of the reconstruction is increased, the centered part becomes more accurate and the dissipation terms become smaller because the jumps in the dissipative parts become smaller. The previous statement holds when the solution is sufficiently smooth. Of course, if the solution is non-smooth, the dissipative contributions increase so as to give the scheme much-needed stabilization. This is how the multidimensional Riemann solver stabilizes the solution.

It is realized that most practitioners would like to see explicit forms for all components of the electric and magnetic fields in eqns. (4.2) and (4.3). Since this might aid in numerical implementation, such information is provided in Appendix C.

We also point out that methods discussed in Balsara and Dumbser [18] ensure that the present advances with the multidimensional Riemann solver also extend to unstructured meshes. In this paper we have dealt with electrodynamics without the use of constitutive relations between the electric field and the electric displacement and, likewise, between the magnetic field and the magnetic induction. There has been considerable recent interest (Barbas and Velarde [24], Ismagilov [45], Munz *et al.* [64]) in developing one-dimensional Riemann solvers for electrodynamics in material media with constitutive relations. These are used in conjunction with finite volume discretizations of electrodynamics. Such discretizations do not preserve the essential constraints that are inherent in Maxwell's equations. The MuSIC Riemann solver approach provides the only natural way of extending these Godunov-type methods so that they do indeed preserve the constraints.

**IV.b) Stability of the Multidimensional Riemann Solver in the First Order Limit**

Since this is a novel approach to electrodynamics, it helps to show that the proposed multidimensional Riemann solver always has a stabilizing effect at first order. To that end, let the integer triplet $(i, j, k)$ denote the center of a zone and let us suitably use half integer indices to denote faces or edges. For example, $(i+1/2, j, k)$ denotes a variable in the x-face and $(i+1/2, j+1/2, k)$ denotes a variable in the z-edge. A uniform Cartesian mesh with zones of size $\Delta x$, $\Delta y$ and $\Delta z$ in the x,y,z-directions is assumed. We can then write

$$E^{z*}_{i+1/2,j+1/2,k} = \overline{E}^{z}_{i+1/2,j+1/2,k} + \frac{1}{2}\left(B^{y}_{i+1,j+1/2,k} - B^{y}_{i,j+1/2,k}\right) - \frac{1}{2}\left(B^{x}_{i+1/2,j+1,k} - B^{x}_{i+1/2,j,k}\right)$$

$$\text{with} \quad \overline{E}^{z}_{i+1/2,j+1/2,k} \equiv \frac{1}{8}\begin{pmatrix} E^{z}_{i,j,k+1/2} + E^{z}_{i,j,k-1/2} + E^{z}_{i+1,j,k+1/2} + E^{z}_{i+1,j,k-1/2} \\ +E^{z}_{i,j+1,k+1/2} + E^{z}_{i,j+1,k-1/2} + E^{z}_{i+1,j+1,k+1/2} + E^{z}_{i+1,j+1,k-1/2} \end{pmatrix} \quad (4.4)$$

where the centered z-component of the electric field is shown with an overbar. In the first order case the centered z-component of the electric field is just an arithmetic average of the eight magnetic field components in the eight z-faces that are closest to the z-edge. The dissipation is carried by the jumps in the x- and y-components of the magnetic field in the x- and y-faces that abut that z-edge. We can then write the first order accurate update equation for the x-component of the magnetic field as

$$\frac{\partial B^{x}_{i+1/2,j,k}}{\partial t} = -\frac{c}{\Delta y}\left(E^{z*}_{i+1/2,j+1/2,k} - E^{z*}_{i+1/2,j-1/2,k}\right) + \frac{c}{\Delta z}\left(E^{y*}_{i+1/2,j,k+1/2} - E^{y*}_{i+1/2,j,k-1/2}\right) \quad (4.5)$$

Substituting eqn. (4.4) and its analogues into eqn. (4.5), and using the discrete divergence-free condition for the magnetic field, we get

$$\frac{\partial B^{x}_{i+1/2,j,k}}{\partial t} = -\frac{c}{\Delta y}\left(\overline{E}^{z*}_{i+1/2,j+1/2,k} - \overline{E}^{z*}_{i+1/2,j-1/2,k}\right) + \frac{c}{\Delta z}\left(\overline{E}^{y*}_{i+1/2,j,k+1/2} - \overline{E}^{y*}_{i+1/2,j,k-1/2}\right)$$
$$+ \frac{c}{2\Delta x}\left(B^{x}_{i+3/2,j,k} - 2B^{x}_{i+1/2,j,k} + B^{x}_{i-1/2,j,k}\right) + \frac{c}{2\Delta y}\left(B^{x}_{i+1/2,j+1,k} - 2B^{x}_{i+1/2,j,k} + B^{x}_{i+1/2,j-1,k}\right) \quad (4.6)$$
$$+ \frac{c}{2\Delta z}\left(B^{x}_{i+1/2,j,k+1} - 2B^{x}_{i+1/2,j,k} + B^{x}_{i+1/2,j,k-1}\right)$$

It is easy to identify the central part in the first line of eqn. (4.6). The remaining lines of eqn. (4.6) show that the dissipative parts are indeed parabolic. Notice too that the diffusion coefficient for the parabolic part in eqn. (4.6) is mesh-dependent and reduces as the zone size decreases – this is consistent with the notion of a first order accurate scheme. Cyclic rotation of the components and the subscripts shows that all components of the magnetic field have dissipative parts that are also parabolic. For the homogeneous system shown in eqn. (4.1), analogous update equations (with parabolic dissipation terms) can be obtained for the time-update of the electric field components. Application of the multidimensional Riemann solver, therefore, stabilizes the update equations for electrodynamics.

The discussion in the present sub-section has focused on the first order case. Please also recall that with increasing accuracy, the central part will be evaluated more accurately and the contribution of the parabolic parts will be reduced when the solution is smooth. This will also be reinforced in the accuracy analysis that we present in Section VII.

**V) Higher Order Reconstruction for the Fluid Variables and Higher Order Treatment of Source Terms**

Two big challenges confront any relativistic code for two-fluid electrodynamics. First, we want the higher order reconstruction of flow variables to preserve physical realizability. This issue is, of course, as relevant at second order as it is for accuracies that are higher than second order. Even non-relativistic flows need to maintain positivity of density and pressure in order to remain physically realizable. But the issue becomes especially severe for relativistic flow because relativistic flows need to remain subluminal. In other words, the flow speed should always remain lower than the speed of light. Sub-section V.a shows how this is achieved. Second, the source terms on the right hand side of the governing equations have to be treated implicitly if large timesteps are to be achieved. While this issue also affects a second order scheme with stiff source terms, it becomes especially severe for a higher order scheme which has to treat the source terms with high order spatial and temporal accuracy. This issue is treated in Sub-section V.b.

**V.a) High Accuracy Subluminal Reconstruction of Relativistic Fluid Variables**

This section follows the innovations described in Balsara and Kim [22] where a method that retains sub-luminal velocities in space and time throughout a timestep is described. Designing a method that retains sub-luminal velocities not just in space but even in time requires the invention of an ADER scheme for time evolution (Titarev and Toro [77], [78], Toro and Titarev [79], Dumbser *et al*. [33], Balsara et al. [10], [11]). Unfortunately, our study of relativistic two-fluid electrodynamics has not advanced so far that we can present an analogous ADER scheme for this PDE system. However, the advance reported here does indeed catalogue a spatial reconstruction strategy that is at least good to fourth order accuracy and also retains sub-luminal flow velocities.

We realize, first off, that in order to ensure positivity of density and pressure we need to carry out the reconstruction in those two primitive variables. This is, of course, done for the positron/proton and the electron fluids. But we also follow Balsara and Kim [22] in realizing that the velocity reconstruction should be carried out for $\gamma \mathbf{v}$ where $\gamma$ is the Lorentz factor and $\mathbf{v}$ is the vector of the three-velocity. If the reconstructed variable $\gamma \mathbf{v}$ is available anywhere within a zone, we are free to define a positive variable $\vartheta \equiv \gamma^2 \mathbf{v}^2$ at any location within the same zone. The magnitude of the three-velocity is then given by $|\mathbf{v}| = \sqrt{\vartheta/(1+\vartheta)}$ which is always smaller than unity, i.e. the three-velocity is always sub-luminal. The Lorentz factor is then given by $\gamma = \sqrt{1+\vartheta}$, which shows that if $\gamma \mathbf{v}$ is given anywhere with a zone, we can obtain the Lorentz factor and the sub-luminal three-velocity with a very small number of float point operations. This reconstruction strategy is applied to the positron/proton fluid's velocity and also to the electron fluid's velocity.

If applied literally, such a reconstruction strategy would work up to second order of accuracy. It is, however, commonly felt among the higher order numerical methods community that in order to obtain third or fourth accuracy, the reconstruction must be carried out in the conserved variables. The work of McCorquodale and Colella [59] is an exception and indeed shows that higher order finite volume reconstruction can be carried out on the primitive variables up to fourth order of accuracy if sufficient care is taken. We very briefly describe this encode-decode algorithm of McCorquodale and Colella [59] because it plays an important role not just in the reconstruction that is documented in this Sub-section but also in the higher order treatment of source terms that is documented in the next Sub-section.

The crucial idea of McCorquodale and Colella is contained in their eqns. (12) and (16) which we draw on here. Let $\bar{\mathbf{U}}_{i,j,k}$ denote any zone-averaged vector of conserved variables in the zone $(i,j,k)$. From that, we can always extract $\mathbf{U}_{i,j,k}$ which gives us the value of the vector of conserved variables exactly at the center of that zone. We define a flattener function $\phi_{i,j,k}$ within that zone such that $\phi_{i,j,k}=1$ for smooth flow and $\phi_{i,j,k}\to 0$ when the flow becomes increasingly non-smooth. The description of such flattener functions for various PDE systems is given in Colella and Woodward [30], Balsara [20] and Balsara and Kim [22]. If the flow is smooth, i.e. if $\phi_{i,j,k}=1$, fourth order accuracy can be ensured. In the fourth order limit we can assert

$$\mathbf{U}_{i,j,k} = \bar{\mathbf{U}}_{i,j,k} - \frac{1}{24}\phi_{i,j,k}\left[\left(\bar{\mathbf{U}}_{i+1,j,k} - 2\bar{\mathbf{U}}_{i,j,k} + \bar{\mathbf{U}}_{i-1,j,k}\right) + \left(\bar{\mathbf{U}}_{i,j+1,k} - 2\bar{\mathbf{U}}_{i,j,k} + \bar{\mathbf{U}}_{i,j-1,k}\right) + \left(\bar{\mathbf{U}}_{i,j,k+1} - 2\bar{\mathbf{U}}_{i,j,k} + \bar{\mathbf{U}}_{i,j,k-1}\right)\right]$$
(5.1)

In other words, when the flow is sufficiently smooth, eqn. (5.1) will give us a fourth order accurate value for $\mathbf{U}_{i,j,k}$, the vector of conserved variables which is defined pointwise at the zone center. Using standard root-solver processes, it is possible to obtain $\mathbf{V}_{i,j,k}$, the vector of primitive variables defined pointwise at the zone center, from the vector $\mathbf{U}_{i,j,k}$. Once $\mathbf{V}_{i,j,k}$ is obtained, we can obtain the zone-averaged vector of primitive variables $\bar{\mathbf{V}}_{i,j,k}$ by using the inverse procedure to eqn. (5.1). We get

$$\bar{\mathbf{V}}_{i,j,k} = \mathbf{V}_{i,j,k} + \frac{1}{24}\phi_{i,j,k}\left[\left(\mathbf{V}_{i+1,j,k} - 2\mathbf{V}_{i,j,k} + \mathbf{V}_{i-1,j,k}\right) + \left(\mathbf{V}_{i,j+1,k} - 2\mathbf{V}_{i,j,k} + \mathbf{V}_{i,j-1,k}\right) + \left(\mathbf{V}_{i,j,k+1} - 2\mathbf{V}_{i,j,k} + \mathbf{V}_{i,j,k-1}\right)\right]$$
(5.2)

In flow regions where fourth order accuracy is not warranted, eqns. (5.1) and (5.2) revert to second order accuracy, which is indeed more stabilizing anyway.

Now that the zone-averaged vector of primitive variables $\bar{\mathbf{V}}_{i,j,k}$ is available in all the zones, we can use that information to carry out multidimensional finite volume WENO or PPM reconstruction within the zone with up to fourth order of accuracy (Colella and Woodward [30], Jiang and Shu [46], Balsara and Shu [9], Dumbser and Käser [34], Balsara *et al.* [10], Colella and Sekora [29], McCorquodale and Colella [59]). Methods from Balsara [20] can always be used to ensure that the density and pressure are positive and methods from Balsara and Kim [22] ensure that the velocity remains sub-luminal. This completes our description of a subluminal reconstruction of relativistic flow variables that can be up to fourth order spatially accurate.

**V.b) Higher Order Treatment of Stiff Source Terms**

The best way to achieve high order accuracy in space and time is to use ADER schemes, as is done in Balsara and Kim [22], because such schemes ensure subluminal evolution of the velocity throughout the course of a timestep. Such methods are not in hand for relativistic two-fluid electrodynamics, which is why we use the Runge-Kutta IMEX methods from Pareschi and Russo [66]; see also Kupka *et al*. [55]. IMEX stands for implicit-explicit, so that the flux terms are treated explicitly while the source terms are treated in implicit fashion. The IMEX methods have the twin advantages of being asymptotically preserving and being stiffly-accurate. Since the equations described in Section II are of relaxation type (Kumar and Mishra [54]), using time update methods that have these dual benefits is indeed a tremendous advantage. The IMEX schemes by Pareschi and Russo are second or third order accurate. Fourth and fifth order accurate IMEX schemes have been designed (Hunsdorfer and Ruuth [44]) but they are too unwieldy for practical usage. We, therefore, restrict attention to second and third order IMEX schemes.

While describing higher order extensions of their IMEX methods, Pareschi and Russo devote an entire section to higher order space-time methods. (Please see Section 4 from Pareschi and Russo [66].) They conclude that finite difference methods are easily extended to high order. Second order finite volume methods are also easily treated; because at second order the finite difference and finite volume methods become identical. However, as noted by Pareschi and Russo, higher order finite volume methods require more care because the source term has to be integrated over the zone with sufficient accuracy. We put the philosophy of McCorquodale and Colella [59] to an entirely different use by showing how to implement higher order finite volume IMEX methods in a fashion that uses computer memory as well as the computer's processor in a maximally efficient fashion.

We first catalogue two very popular Runge-Kutta IMEX schemes from Pareschi and Russo [66] in their finite volume forms. The first scheme is the second order accurate IMEX-SSP2(3,2,2) stiffly-accurate scheme given by

$$\bar{\mathbf{U}}^{(1)} = \bar{\mathbf{U}}^n + \frac{\Delta t}{2} \overline{R(\mathbf{U}^{(1)})} \tag{5.3}$$

$$\bar{\mathbf{U}}^{(2)} = \bar{\mathbf{U}}^n - \frac{\Delta t}{2} \overline{R(\mathbf{U}^{(1)})} + \frac{\Delta t}{2} \overline{R(\mathbf{U}^{(2)})} \tag{5.4}$$

$$\bar{\mathbf{U}}^{(3)} = \bar{\mathbf{U}}^n + \Delta t \overline{L(\mathbf{U}^{(2)})} + \frac{\Delta t}{2} \overline{R(\mathbf{U}^{(2)})} + \frac{\Delta t}{2} \overline{R(\mathbf{U}^{(3)})} \tag{5.5}$$

$$\bar{\mathbf{U}}^{n+1} = \bar{\mathbf{U}}^n + \frac{\Delta t}{2} \overline{L(\mathbf{U}^{(2)})} + \frac{\Delta t}{2} \overline{L(\mathbf{U}^{(3)})} + \frac{\Delta t}{2} \overline{R(\mathbf{U}^{(2)})} + \frac{\Delta t}{2} \overline{R(\mathbf{U}^{(3)})} \tag{5.6}$$

Notice that this timestepping scheme has only two flux evaluations and is, therefore, only second order accurate in time even if it makes multiple evaluations of the source terms in order to be asymptotically stable. Without the source terms, it can be rewritten in a format that exactly reproduces the SSP-RK2 scheme of Shu and Osher [70]. Here the subscripts of $(i, j, k)$ for a zone have been dropped because the source terms are always treated on a zone-by-zone basis. Just to take one example, focus on the term $\overline{L(\mathbf{U}^{(2)})}$ in eqn. (5.5). It refers to the contribution of the flux terms (evaluated in a finite volume sense) to the time update of $\bar{\mathbf{U}}^{(3)}$. This flux term is averaged over the entire zone in a finite volume sense. The term $\overline{R(\mathbf{U}^{(3)})}$ in eqn. (5.5) refers to a spatially high order, and fully implicit, treatment of the source terms. The higher order treatment of such source terms were problematical for Pareschi and Russo [66]. We will soon show that such a calculation can be arranged even at high order, and even in three dimensions, in a form that is economical in its use of computer memory and computer processor resources.

Our second scheme is the third order accurate IMEX-SSP3(4,3,3) scheme given by

$$\bar{\mathbf{U}}^{(1)} = \bar{\mathbf{U}}^n + \alpha \Delta t \overline{R(\mathbf{U}^{(1)})} \tag{5.7}$$

$$\overline{\mathbf{U}}^{(2)} = \overline{\mathbf{U}}^n - \alpha\,\Delta t\,\overline{R(\mathbf{U}^{(1)})} + \alpha\,\Delta t\,\overline{R(\mathbf{U}^{(2)})} \tag{5.8}$$

$$\overline{\mathbf{U}}^{(3)} = \overline{\mathbf{U}}^n + \Delta t\,\overline{L(\mathbf{U}^{(2)})} + (1-\alpha)\Delta t\,\overline{R(\mathbf{U}^{(2)})} + \alpha\,\Delta t\,\overline{R(\mathbf{U}^{(3)})} \tag{5.9}$$

$$\overline{\mathbf{U}}^{(4)} = \overline{\mathbf{U}}^n + \frac{\Delta t}{4}\overline{L(\mathbf{U}^{(2)})} + \frac{\Delta t}{4}\overline{L(\mathbf{U}^{(3)})} + \beta\,\Delta t\,\overline{R(\mathbf{U}^{(1)})} + \eta\,\Delta t\,\overline{R(\mathbf{U}^{(2)})}$$
$$+ \left(\frac{1}{2}-\beta-\eta-\alpha\right)\Delta t\,\overline{R(\mathbf{U}^{(3)})} + \alpha\,\Delta t\,\overline{R(\mathbf{U}^{(4)})} \tag{5.10}$$

$$\overline{\mathbf{U}}^{n+1} = \overline{\mathbf{U}}^n + \frac{\Delta t}{6}\overline{L(\mathbf{U}^{(2)})} + \frac{\Delta t}{6}\overline{L(\mathbf{U}^{(3)})} + \frac{2\Delta t}{3}\overline{L(\mathbf{U}^{(4)})}$$
$$+ \Delta t\frac{1}{\varepsilon}\left[\frac{1}{6}\overline{R(\mathbf{U}^{(2)})} + \frac{1}{6}\overline{R(\mathbf{U}^{(3)})} + \frac{2}{3}\overline{R(\mathbf{U}^{(4)})}\right] \tag{5.11}$$

The coefficients in the above equations are given by

$$\alpha = 0.24169426078821 \quad ; \quad \beta = 0.06042356519705 \quad ; \quad \eta = 0.12915286960590 \tag{5.12}$$

This timestepping scheme has three flux evaluations and is, therefore, only third order accurate in time. Without the source terms, it can be rewritten in a format that exactly reproduces the SSP-RK3 scheme of Shu and Osher [70]. It is the source terms $\overline{R(\mathbf{U}^{(1)})}$ through $\overline{R(\mathbf{U}^{(4)})}$ that have to be evaluated with at least third order spatial accuracy if a scheme that is third order accurate in space and time is to be achieved. Likewise, if the same source terms are evaluated with fourth order of spatial accuracy, we will get a scheme that is third order in time but fourth order in space. The goal in this Sub-section is to arrange the calculation efficiently so that the target spatial and temporal accuracy is achieved.

Within each zone we pick six nodal points that are collocated at the face-centers of that zone. The values at those nodal points are evaluated from *within* the zone using the third or fourth order accurate finite volume reconstruction of $\overline{\mathbf{U}}^n$ that has been carried out within the zone in question. (In practice, the third or fourth order reconstruction that has been carried out on the primitive variables can also be used at those six nodal points.) Therefore, within each zone we have a set $\{\mathbf{U}_m^n : m=1,...,6\}$ of conserved variables at each of the six nodal points. The flux terms provide the same amount of temporal update to all those six terms because this is a finite volume scheme. However, we define six source terms within each zone that are located at each of the six nodal points. These six source terms are defined for each stage in the update described by eqns. (5.7) to (5.11). To take eqn. (5.10) as an example, it will be used to find the six elements of the set $\{R(\mathbf{U}_m^{(4)}) : m = 1,...,6\}$ in a fashion that will soon be described. The sets of source terms given by $\{R(\mathbf{U}_m^{(3)}) : m = 1,...,6\}$, $\{R(\mathbf{U}_m^{(2)}) : m = 1,...,6\}$ and $\{R(\mathbf{U}_m^{(1)}) : m = 1,...,6\}$ have already been built up and stored within each zone from the stages given by eqns. (5.7), (5.8) and (5.9) respectively. The sole purpose of the implicit eqn. (5.10) is, therefore, to build the set of conserved variables $\{\mathbf{U}_m^{(4)} : m = 1,...,6\}$ and the corresponding set of source terms $\{R(\mathbf{U}_m^{(4)}) : m = 1,...,6\}$. Notice that all the conserved variables that are found implicitly within a zone, as well as their corresponding source terms, are collocated at the six nodal points within the zone which are always evaluated from inside the zone during the course of the timestep. Thus eqn. (5.10) can be viewed as six nodal equations that have to be solved implicitly. Any one of those equations is given by

$$\mathbf{U}_m^{(4)} - \alpha \, \Delta t \, R(\mathbf{U}_m^{(4)}) = \mathbf{U}_m^n + \Delta t \left\{ \frac{1}{4} \overline{L(\mathbf{U}^{(2)})} + \frac{1}{4} \overline{L(\mathbf{U}^{(3)})} \right\}$$
$$+ \Delta t \left\{ \beta \, R(\mathbf{U}_m^{(1)}) + \eta \, R(\mathbf{U}_m^{(2)}) + \left(\frac{1}{2} - \beta - \eta - \alpha\right) R(\mathbf{U}_m^{(3)}) \right\} \; ; \; m = 1,...,6$$
(5.13)

The left hand side of eqn. (5.13) has to be solved implicitly within each zone for $\{\mathbf{U}_m^{(4)} : m = 1,...,6\}$. Newton iteration applied at each of the six nodal points works well and enables the conserved variables to converge rapidly. From these converged, conserved variables at the six nodal points, we can evaluate $\{R(\mathbf{U}_m^{(4)}) : m = 1,...,6\}$. Notice that the flux terms, $\overline{L(\mathbf{U}^{(2)})}$ and $\overline{L(\mathbf{U}^{(3)})}$, provide the same contribution to each of the nodal points. For each stage in the Runge-Kutta update they can, therefore, be clubbed into a single term that is shown by the first curly brackets in eqn. (5.13). However, the three source terms from the previous stages on the right hand side of eqn. (5.13) (i.e., $R(\mathbf{U}_m^{(1)})$, $R(\mathbf{U}_m^{(2)})$ and $R(\mathbf{U}_m^{(3)})$) provide different amounts of contribution at the different nodal points within a zone. For each stage in the Runge-Kutta update they can, therefore, be clubbed into a single term that is shown by the second curly brackets in eqn. (5.13). By comparing eqn. (5.13) to all the other stages in the Runge-Kutta update, we realize that the $p^{th}$ stage can always be written in the general form

$$\mathbf{U}_m^{(p)} - \alpha \, \Delta t \, R\left(\mathbf{U}_m^{(p)}\right) = \mathbf{U}_m^n + \Delta t \left\{\text{Flux terms for } p^{th} \text{ stage; same for all nodes within a zone.}\right\}$$
$$+ \Delta t \left\{\text{Previous source terms for } p^{th} \text{ stage; specific for each node in a zone.}\right\} \quad (5.14)$$
$$; \quad m = 1, \ldots, 6$$

Once eqn. (5.13) has been solved at each of the six nodal points, we can obtain the zone average from this stage, i.e. $\overline{\mathbf{U}}^{(4)}$, as

$$\overline{\mathbf{U}}^{(4)} = \frac{1}{6}\left(\mathbf{U}_1^{(4)} + \mathbf{U}_2^{(4)} + \mathbf{U}_3^{(4)} + \mathbf{U}_4^{(4)} + \mathbf{U}_5^{(4)} + \mathbf{U}_6^{(4)}\right) \quad (5.15)$$

Once eqn. (5.13) has been solved at each of the six nodal points, the six source terms $\left\{R\left(\mathbf{U}_m^{(4)}\right) : m = 1, \ldots, 6\right\}$ should also be stored for each zone. The zone average $\overline{\mathbf{U}}^{(4)}$ can be used for the reconstruction and flux evaluation that builds the term $\overline{L\left(\mathbf{U}^{(4)}\right)}$ in eqn. (5.11). In that fashion, the zone averages from each of the Runge-Kutta stages in eqns. (5.9) through (5.10) contributes to the flux evaluation in the Runge-Kutta stages that follow them. The generic form for any IMEX Runge-Kutta stage at high order is catalogued in eqn. (5.14). In practical terms, the subroutine in the code that updates the stages should have the general form given in eqn. (5.14). This ensures that the flux evaluation is explicit in time. The quadrature in eqn. (5.15) is spatially fourth order accurate and can be applied to each of the stages. Thus the method will be spatially fourth order accurate if the rest of the scheme is spatially fourth order accurate. If the rest of the scheme is spatially third order accurate, eqn. (5.15) will yield a spatially third order accurate scheme. The temporal accuracy is limited to third order in time for IMEX-SSP3(4,3,3).

While RKDG schemes are not the topic of this investigation, the present method also extends to such schemes. In those schemes, the Runge-Kutta method is used to update not just the flow variable but also its higher moments. These higher moments constitute the modal representation of the RKDG scheme. In that case, we should pick a number of nodes within each zone that is exactly commensurate with the number of modes being updated. The nodes should also be picked at special locations so that one can make a modal to nodal transcription (Balsara *et al*. [10]). At each of the nodes, equations that are analogous to eqn. (5.13) can be solved. As long as the modal to nodal transcription is invertible, we can subsequently find the modal variables at the updated time. This paragraph provides a quick and useful extension of the ideas in this Sub-section for treating stiff source terms in RKDG schemes.

While the method is stabilized by making the source terms implicit, it is important to point out that the convergence is achieved via a Newton iteration. I.e., the equations are linearized in the Newton iteration step. Eventually, the saliency of the linearization will limit the timesteps that can be taken by the method. We provide a physics-based criterion for controlling the timestep. Realize that eqn. (2.4) evolves the mass density while eqn. (2.15) evolves the charge density. When there is substantial amount of charge separation in a plasma, the Coulomb forces grow to large values. Numerically, a similar phenomenon might occur. Thus our suggestion consists of monitoring the ratio

$$\frac{\min\left(\frac{m_e}{|q_e|}, \frac{m_p}{q_p}\right)\left(\frac{q_e}{m_e}\gamma_e\rho_e + \frac{q_p}{m_p}\gamma_p\rho_p\right)}{\left(\gamma_e\rho_e + \gamma_p\rho_p\right)} \quad (5.16)$$

When the above ratio becomes larger than about $10^{-4}$ on the mesh, our observation has been that numerically induced instabilities may set in. In such a situation, the timestep should be reduced till the above ratio is brought within limits.

## VI) Pointwise Description of the Algorithm

To make the algorithm more accessible, we describe it in pointwise fashion below. We describe it in two stages. The next paragraph describes just the algorithm for updating the electromagnetic field when the two-fluid terms are absent. The subsequent pointwise description describes the algorithm in the presence of the relativistic two-fluid terms.

It is useful to describe just the electromagnetic part of the algorithm because it can be used as a stand-alone solver for the Maxwell equations. The relativistic two-fluid equations then drop out of consideration. In that situation, one is exclusively dealing with eqns. (2.9) to (2.12) along with the evolutionary equation for the charge density, eqn. (2.16). In that case, the current density is constitutively related to the electric field. Eqn. (2.16), therefore, provides an evolutionary equation for the zone-centered charge density. The charge density is reconstructed using any higher order finite volume reconstruction procedure like WENO or PPM. The moments of the facial components of the electric and magnetic fields are then reconstructed using the same reconstruction strategies. Using Section III, we can obtain the reconstructed electric and magnetic fields within each zone. The reconstructed electric and magnetic fields are then evaluated at each edge. For higher order schemes, multiple quadrature points may be used within each edge. The multidimensional Riemann solver is invoked at each edge. This gives us the higher order edge-averaged electric and magnetic fields. These are then used, along with the facially-centered current densities, to update the facial electric and magnetic fields using a Stokes law-type update. This completes our description of a stand-alone solver for the Maxwell equations that is based on the methods described in this paper.

We now describe the algorithm for relativistic two-fluid electrodynamics in pointwise form. This is done for a single stage in the multi-stage IMEX-based Runge-Kutta schemes described in Sub-section V.b. The computational tasks within each stage are as follows:

**1)** Using a second order procedure, evaluate the flattener function, $\phi_{i,j,k}$, within each zone. The flattener does not need to be evaluated with high order of accuracy.

**2)** Use eqn. (3.6) to evaluate the volume-averaged charge density within each zone. Using any traditional finite-volume reconstruction strategy, obtain the higher moments of the charge density that are to be used in eqns. (3.4), (A.4) or (B.4).

**3)** Using any higher order reconstruction strategy, obtain the moments of the electric and magnetic field components within each face. With these facial moments for the electric field in hand, and with the higher moments of the charge density in hand, we use the methods from Section III to obtain the consistent reconstructed electric field within each zone; i.e., this electric field will be consistent with Gauss' law. Similar methods from Balsara [8] can be used to obtain a divergence-free reconstruction of the magnetic field within each zone.

**4)** Eqns. (5.1) and (5.2) are used to obtain volume-averaged primitive variables. (Please note that the reconstructed electric and magnetic fields are needed in order to obtain the primitive variables from the conserved variables. For this reason, we need to reconstruct the electric and magnetic fields before we can obtain the zone-averaged primitive variables.) Standard finite volume-based reconstruction can be used on the primitive variables to obtain the reconstructed flow variables for the positively- or negatively-charged fluids. The velocity reconstruction is done using the $\gamma\mathbf{v}$ variable, as described in Sub-section V.a. This ensures that the reconstructed three-velocity remains sub-luminal within each zone. Also use the method from Balsara [20] to ensure that the reconstructed density and pressure are positive throughout each zone.

**5)** By this point, all the reconstructed flow variables for the positively- and negatively-charged fluids are available at all points on the mesh. Likewise, the reconstructed electric and magnetic fields are available at all points on the mesh. We now wish to obtain properly upwinded numerical fluxes. Likewise, we wish to obtain multidimensionally upwinded electric and magnetic fields within each edge. Using a suitable quadrature strategy within each face, obtain the facially-averaged numerical flux within that face. This is done by invoking one-dimensional Riemann solvers at each of the quadrature points. Similarly, invoke the multidimensional Riemann solver described in Sub-section IV.a at a suitable number of quadrature points within each edge. Use the output from the multidimensional Riemann solver to obtain edge-averaged electric and magnetic fields.

**6)** Using a Stokes-law type of update strategy, update the facial electric and magnetic fields. The numerical fluxes from eqn. (2.15) also provide the facially-averaged current densities which contribute to the update of the facial electric fields. It is important to update the facial electric and magnetic fields before the implicit part of the IMEX update of the zone-centered variables.

**7)** Realize that the entire remaining task consists of updating equations that schematically look like eqn. (5.14). Reconstruct the updated electric and magnetic fields a second time within this stage. At each of the six quadrature points, eqn. (5.14) constitutes an implicit solve for the ten fluid variables for the positively- and negatively-charged fluids. The updated and reconstructed electric and magnetic fields are needed on the left hand side of eqn. (5.14); however, they do not need any further update.

**8)** Once the conserved variable is available at each of the six quadrature points within each zone, use eqn. (5.15) to obtain the zone-averaged conserved variables at the end of this stage. Use the conserved variables to also build the source terms at the six quadrature points and store them for use within the next stage.

This completes our pointwise description of the algorithm.

Applications-oriented practitioners are more accustomed to traditional higher order Godunov schemes with their zone-centered flow variables and face-centered fluxes. Such practitioners might perhaps find it difficult to think in terms of the face-centered electric and magnetic fields as being the primary variables of the method. Furthermore, the face-centered electric fields are updated using edge-centered magnetic fields. Likewise, the face-centered magnetic fields are updated using edge-centered electric fields. Because of this staggering, practitioners with a background in higher order Godunov schemes might still appreciate some additional helpful advice on implementation; we do that here. We would advise them to first start with declaring data with the appropriate collocation, i.e. whether it is face-centered or edge-centered. We would then advise them to start with the second order scheme. It is easiest to focus on second order schemes and, once a second order scheme is implemented, higher order schemes follow quite naturally. The following two paragraphs make some helpful comments about the constraint-preserving reconstruction and the edge-centered update.

Visit (loop over) each face and look at its neighboring faces in either direction. Use the electric and magnetic field components in the neighboring faces to obtain the slopes within the face of interest. Store those slopes at each facial location. Now make sure that the zone-centered gradients associated with the charge density are built using any reasonable limiting procedure. Then visit (loop over) each zone and use the field components and their slopes, which reside in the faces of that zone, to carry out the reconstruction that is described in Section III. This results in a constraint-preserving reconstruction within each zone.

To evaluate the edge-centered electric and magnetic fields, visit each edge and obtain the four electric and magnetic fields from the four zones that surround it. Please see Appendix C to realize that only a few such electric and magnetic field components are needed. The reconstruction from the previous paragraph is very useful in providing the necessary fields with high order of accuracy. Appendix C then provides the explicit formulae for obtaining the edge-centered electric and magnetic fields.

We compared the CPU time (per timestep) of a higher order Godunov scheme which evolves the sixteen zone-centered variables to the method presented here which evolved ten zone-centered variables in higher order Godunov fashion along with the six face-centered fields in constraint-preserving fashion. The method presented here adds less than than 7% more CPU time to the overall cost of the scheme. Realize too that a traditional higher order Godunov method with zone-centered electric and magnetic fields would also require the application of the same limiters etc. to achieve higher order accuracy. The constraint-preserving reconstruction does not add to that cost. Also realize that each of the edge-centered electric and magnetic field evaluations in Appendix C requires only a few float point operations. This is why the additional cost, over and above that of a traditional higher order Godunov scheme, is indeed minimal. It is worth mentioning that in Sub-Section VIII.b we show an example where a traditional higher order Godunov scheme can build up divergence in the magnetic field and, therefore, prompt the formation of spurious flow features. If the problem does not entail periodic geometry, one may perhaps be able to use the GLM method of Munz *et al*. [64] to propagate away the spurious features. In periodic geometry, the spurious divergence-related effects would continue to stay on the mesh when the GLM method is used. Note too that the GLM method will involve the addition of extra flow variables associated with the Lagrange multipliers, which add to the computational cost. Furthermore, at stagnation points, or for problems with periodic geometries, it may not always be possible to remove the build-up of spurious divergence. The method presented in this paper always maintains the requisite consistency. We feel that the tiny additional cost (i.e. a cost increase of less than 7%) is indeed a very small price to pay for a truly consistent evolution of electric and magnetic fields in all possible geometries and for all flow configurations.

## VII) Accuracy Analysis

Three accuracy analyses are presented. The first deals with the propagation of a plane polarized electromagnetic wave in vacuum, the second deals with the propagation of a torsional

Alfven wave, the third deals with the propagation of a circularly polarized electromagnetic wave in a plasma. The physics of the second and third test problems is such that they are susceptible to a slight parametric instability. Consequently, the time over which these test problems can be run without the instability cropping up is limited. The first test problem is not susceptible to such a limitation. The first test does not involve stiff source terms so we use ordinary SSP-RK methods; the next two tests involve source terms which call for IMEX-based Runge-Kutta methods.

For the first accuracy test (in Sub-section VII.a) we show that second, third and fourth orders of accuracy can be achieved on two- and three-dimensional Cartesian meshes. Please note though that the methods presented here do extend naturally to even higher orders and to more complicated mesh geometries (Balsara and Dumbser [18]). The spatial reconstruction used here is a finite volume-based, weighted essentially-non oscillatory (WENO) scheme (Balsara *et al*. [10], [11]). More information on WENO methods can also be found (Shu and Osher [70], [71], Jiang and Shu [46], Balsara and Shu [9], Dumbser and Käser [35]). The temporal evolution is based on strong-stability preserving Runge-Kutta schemes (Shu and Osher [70], [71], Shu [72], Spiteri and Ruuth [73], [74], Gottlieb [39], Gottlieb, Shu and Tadmor [40], Gottlieb, Ketcheson and Shu [41]). Specifically, we use the second and third order accurate SSP-RK2 and SSP-RK3 schemes from Shu and Osher [70] and the fourth order accurate SSP-RK(5,4) scheme (Spiteri and Ruuth [73], [74], Gottlieb [39]).

For the second and third accuracy tests (in Sub-sections VII.b and VII.c respectively), we use IMEX Runge-Kutta methods from Pareschi and Russo [66]. The spatial reconstruction is still based on WENO schemes. The scheme with second order spatial accuracy is coupled with the second order IMEX scheme that is catalogued in eqns. (5.3) to (5.6). The schemes with third and fourth order of spatial accuracy are coupled with the third order IMEX scheme that is catalogued in eqns. (5.7) to (5.12). As a result, their overall accuracy will be limited to third order. However, we will show that the schemes with fourth order spatial accuracy provide a considerable overall advantage. Fourth and fifth order IMEX schemes from Hunsdorfer [44] were deemed to be too difficult to see practical implementation in production codes. ADER methods (Titarev & Toro [77], [78], Toro & Titarev [79], Schwartzkopff *et al*. [68], Dumbser *et al*. [33], [34], Balsara *et al*. [10] [11], Loubere et al. [56]) would be very beneficial here, especially because they could enable strong coupling between the source terms in innovative ways. However, this innovation requires time and deserves a paper of its own. It will, therefore, be the topic of a subsequent paper.

## VII.a) Three Dimensional Plane Polarized Electromagnetic Wave in Vacuum

This test problem consists of a plane polarized electromagnetic wave propagating in a vacuum along the diagonal of a three dimensional Cartesian mesh spanning $[-0.5, 0.5]^3$. Periodic boundary conditions are enforced. The easiest way to set up a divergence-free electric and magnetic field is via a vector potential approach. We define the vector potential for the magnetic field to be $\mathbf{A}(x, y, z, t)$ so that the magnetic field is given by $\mathbf{B}(x, y, z, t) = -\nabla \times \mathbf{A}(x, y, z, t)$. Likewise, we define the vector potential for the electric field to be $\mathbf{C}(x, y, z, t)$ so that the electric field itself is given by $\mathbf{E}(x, y, z, t) = -\nabla \times \mathbf{C}(x, y, z, t)$. The electric and magnetic fields are given by

$$\mathbf{E}(x, y, z, t) = \cos\left[2\pi\left(x + y + z - \sqrt{3}t\right)\right]\left(-\frac{1}{\sqrt{2}}\hat{y} + \frac{1}{\sqrt{2}}\hat{z}\right)$$

$$\mathbf{B}(x, y, z, t) = \cos\left[2\pi\left(x + y + z - \sqrt{3}t\right)\right]\left(\sqrt{\frac{2}{3}}\hat{x} - \frac{1}{\sqrt{6}}\hat{y} - \frac{1}{\sqrt{6}}\hat{z}\right)$$

and their vector potentials are given by

$$\mathbf{C}(x, y, z, t) = \frac{1}{2\pi\sqrt{3}}\sin\left[2\pi\left(x + y + z - \sqrt{3}t\right)\right]\left(\sqrt{\frac{2}{3}}\hat{x} - \frac{1}{\sqrt{6}}\hat{y} - \frac{1}{\sqrt{6}}\hat{z}\right)$$

$$\mathbf{A}(x, y, z, t) = \frac{-1}{2\pi\sqrt{3}}\sin\left[2\pi\left(x + y + z - \sqrt{3}t\right)\right]\left(-\frac{1}{\sqrt{2}}\hat{y} + \frac{1}{\sqrt{2}}\hat{z}\right)$$

With these analytical forms in hand, it is possible to evaluate the accuracy of the solution at any later time if it is set up correctly at the initial time on the mesh. The problem was run to a time of unity on the computational mesh.

Table I shows the accuracy analysis for the second order scheme. It used the piecewise linear parts from a centered r=3 WENO reconstruction along with an SSP-RK2 time-integration. The errors and accuracy in the y-component of the electric and magnetic fields are shown at the last time point in the simulation. Because the SSP-RK2 time integration does not have a very high stability bound, we ran this simulation with a CFL of 0.3. We see that the scheme meets its designed second order accuracy. Table II shows the accuracy analysis for the third order scheme. It used the full centered r=3 WENO reconstruction along with an SSP-RK3 time-integration. We ran this simulation with a CFL of 0.45. We see that the scheme is indeed third order accurate. Table III shows the accuracy analysis for the fourth order scheme. It used the full centered r=4 WENO reconstruction along with an SSP-RK(5,4) time-integration. We ran this simulation with a CFL of 0.45. We see that the scheme is indeed fourth order accurate. Our practical experience has been that SSP-RK3 and SSP-RK(5,4) have a substantially superior timestep stability compared to SSP-RK2 for this class of problem. As a result, we would make the practical suggestion that SSP-RK3 should be used even with spatially second order codes, even if that was not the choice that we made here.

By scanning Tables I, II and III we also see that at the fixed resolution the higher order schemes yield vastly better accuracies. On a per timestep basis, the third order scheme has a computational complexity that is 3.8 times the second order scheme. The fourth order scheme has a computational complexity that is 3.8 times the third order scheme. Also consider that the third and fourth order schemes can take timesteps that are 50% larger than the second order scheme. By comparing the accuracies at some of the higher resolutions in Tables I, II and III we see that the improvement in accuracy with increasing order is well in excess of the increase in CPU time. This makes a very good case for designing higher order schemes for computational electrodynamics that are based on the algorithmic elements that are catalogued in this paper. Our practical experience has been that CPUs with larger caches fare a lot better with higher order schemes. The relative timings presented in this paragraph are based on using an Opteron CPU with a middle-of-the-line cache size.

**Table I shows the accuracy analysis for the second order scheme for the propagation of an electromagnetic wave in vacuum. A CFL of 0.3 was used. The errors and accuracy in the y-component of the electric and magnetic fields are shown.**

| zones | Ey $L_1$ error | Ey $L_1$ accuracy | Ey $L_{inf}$ error | Ey $L_{inf}$ accuracy |

| $16^3$ | 0.1655 | | 0.2561 | |
| $32^3$ | 4.6523E-2 | 1.83 | 7.2785E-2 | 1.81 |
| $64^3$ | 1.1909E-2 | 1.96 | 1.8714E-2 | 1.96 |
| $128^3$ | 2.9880E-3 | 1.99 | 4.6927E-3 | 1.99 |
| zones | By $L_1$ error | By $L_1$ accuracy | By $L_{inf}$ error | By $L_{inf}$ accuracy |
| $16^3$ | 0.09560 | | 0.1478 | |
| $32^3$ | 2.6860E-2 | 1.83 | 4.2022E-2 | 1.81 |
| $64^3$ | 6.8758E-3 | 1.96 | 1.0804E-2 | 1.96 |
| $128^3$ | 1.7251E-3 | 1.99 | 2.7093E-3 | 1.99 |

**Table II shows the accuracy analysis for the third order scheme for the propagation of an electromagnetic wave in vacuum. A CFL of 0.45 was used. The errors and accuracy in the y-component of the electric and magnetic fields are shown.**

| zones | Ey $L_1$ error | Ey $L_1$ accuracy | Ey $L_{inf}$ error | Ey $L_{inf}$ accuracy |
|---|---|---|---|---|
| $16^3$ | 4.5310E-2 | | 6.6467E-2 | |
| $32^3$ | 5.9277E-3 | 2.93 | 9.1531E-3 | 2.86 |
| $64^3$ | 7.3790E-4 | 3.00 | 1.1361E-3 | 3.01 |
| $128^3$ | 9.1531E-5 | 3.01 | 1.4181E-4 | 3.00 |
| zones | By $L_1$ error | By $L_1$ accuracy | By $L_{inf}$ error | By $L_{inf}$ accuracy |
| $16^3$ | 2.6159E-2 | | 3.8375E-2 | |
| $32^3$ | 3.4223E-3 | 2.93 | 5.2845E-3 | 2.86 |
| $64^3$ | 4.2603E-4 | 3.00 | 6.5597E-4 | 3.01 |
| $128^3$ | 5.2845E-5 | 3.01 | 8.1875E-5 | 3.00 |

**Table III shows the accuracy analysis for the fourth order scheme for the propagation of an electromagnetic wave in vacuum. A CFL of 0.45 was used. The errors and accuracy in the y-component of the electric and magnetic fields are shown.**

| zones | Ey $L_1$ error | Ey $L_1$ accuracy | Ey $L_{inf}$ error | Ey $L_{inf}$ accuracy |
|---|---|---|---|---|

| zones | By $L_1$ error | By $L_1$ accuracy | By $L_{inf}$ error | By $L_{inf}$ accuracy |
|---|---|---|---|---|
| $16^3$ | 2.8801E-3 | | 3.9768E-3 | |
| $32^3$ | 1.1800E-4 | 4.60 | 1.6803E-4 | 4.56 |
| $64^3$ | 6.2334E-6 | 4.24 | 9.6550E-6 | 4.12 |
| $128^3$ | 3.7060E-7 | 4.07 | 5.8010E-7 | 4.05 |
| zones | By $L_1$ error | By $L_1$ accuracy | By $L_{inf}$ error | By $L_{inf}$ accuracy |
| $16^3$ | 1.6628E-3 | | 2.2960E-3 | |
| $32^3$ | 6.8131E-5 | 4.60 | 9.7013E-5 | 4.56 |
| $64^3$ | 3.5988E-6 | 4.24 | 5.5743E-6 | 4.12 |
| $128^3$ | 2.1397E-7 | 4.07 | 3.3492E-7 | 4.05 |

**VII.b) Two Dimensional Circularly Polarized Alfven Wave in Relativistic Two-Fluid Plasma**

The system of equations from Section II admits Alfven waves and electromagnetic waves. For accuracy analysis of finite amplitude Alfven waves, it helps to use a circularly polarized Alfven wave or a circularly polarized electromagnetic wave (Kennel and Pellat [47]). This ensures that the electromagnetic energy and pressure are constant throughout the wave. For that reason, we use a circularly polarized Alfven wave for this accuracy analysis. The dispersion relation is a quartic which has to be solved numerically. Furthermore, the Lorentz factor for the finite amplitude Alfven waves depends on the amplitude of the fluctuation. In that sense, the situation is rather different from ideal non-relativistic MHD where the amplitude of the Alfven wave can have any value. Our goal in this paper is not to catalogue the solution procedure for solving this quartic. Instead we simply give the parameters that were used for the set up in this Sub-section and the next one.

We should also mention that such finite amplitude circularly polarized Alfven waves and electromagnetic waves are susceptible to a parametric instability in classical single fluid and two-fluid MHD (Goldstein [38], Mima and Nishikawa [62] Terasawa *et al*. [76], Wong and Goldstein [82]) so that a similar parametric instability will also be present in relativistic two-fluid electrodynamics. An analogous parametric instability has been found for relativistic pair plasmas (Matsukiyo and Hada [58]). For that reason, the test problem presented here is an imperfect test problem. The test problem shows its deficiency for high order schemes on large meshes. This is because the numerical dissipation becomes very small on large meshes when high order schemes are used, which gives the instability a chance to manifest itself. As a result, this test problem should be used cautiously.

In this Sub-section and the next, we start with a two-dimensional periodic computational domain in the xy-plane that is twice as long in the x-direction as in the y-direction. A smooth, sinusoidal wave is set up with wavelengths in the x- and y-directions that are equal to the dimensions of the computational domain. The density, pressure and Lorentz factors are constant in each of the two fluids; though each of the two different fluids usually have different values for these quantities. We take the lab frame density in the positive and negative fluids to be unity. The Lorentz factors, $\gamma_p$ and $\gamma_e$, in the two fluids then depend on the amplitude of the wave and play a role in setting the rest frame density of the two fluids. Consequently, we set the rest frame densities in the two fluids to be $\rho_p = 1/\gamma_p$ and $\rho_e = 1/\gamma_e$. The pressures in the two fluids are given by $P_p = 0.01/\gamma_p$ and $P_e = 0.01/\gamma_e$. The polytropic index in both fluids is taken to be $\Gamma = 4/3$. The resistivity terms were assumed to be zero.

A uniform, unperturbed magnetic field is initialized along the diagonal of the computational domain. The unperturbed electric field is zero. To the unperturbed magnetic field we add a transverse finite amplitude perturbation that corresponds to the Alfven wave or the electromagnetic wave. As a result, the final magnetic field is given by

$$\mathbf{B}(x,y,z,t) = B_0 \cos\theta \hat{x} + B_0 \sin\theta \hat{y} + \tilde{B}\left(-\sin\theta \hat{x} + \cos\theta \hat{y}\right)\cos\left(k_x x + k_y y - \omega t\right) - \tilde{B}\,\hat{z}\sin\left(k_x x + k_y y - \omega t\right)$$

Likewise, the final electric field is given by

$$\mathbf{E}(x,y,z,t) = \tilde{E}\left(\sin\theta \hat{x} - \cos\theta \hat{y}\right)\sin\left(k_x x + k_y y - \omega t\right) - \tilde{E}\,\hat{z}\cos\left(k_x x + k_y y - \omega t\right)$$

We use $\theta = \tan^{-1}(1/2)$. Therefore, specifying the unperturbed magnetic field with the coefficient $B_0$, and the perturbation coefficients $\tilde{B}$ and $\tilde{E}$, fully specifies the structure of the magnetic and electric fields. We always keep $k_y = 2k_x$ so that specifying the length of the computational domain in the x-direction fully specifies the size of the computational domain as well as the wave numbers in either direction. For example, $k_x = 2\pi/\lambda_x$, and we will specify the wavelenth $\lambda_x$ in the x-direction for the different test problems used here. A further specification of the angular frequency "ω" then gives us the structure of the sinusoidal wave and its variation in time. The unperturbed velocity is zero in both fluids. The final perturbed three velocity for the positive fluid is given by

$$\mathbf{v}_p(x,y,z,t) = \tilde{v}_p\left(-\sin\theta\hat{x} + \cos\theta\hat{y}\right)\cos\left(k_x x + k_y y - \omega t\right) - \tilde{v}_p\hat{z}\sin\left(k_x x + k_y y - \omega t\right)$$

The final perturbed three velocity for the negative fluid is given by

$$\mathbf{v}_e(x,y,z,t) = \tilde{v}_e\left(-\sin\theta\hat{x} + \cos\theta\hat{y}\right)\cos\left(k_x x + k_y y - \omega t\right) - \tilde{v}_e\hat{z}\sin\left(k_x x + k_y y - \omega t\right)$$

Specifying the perturbed velocity amplitudes $\tilde{v}_p$ and $\tilde{v}_e$ also specifies the Lorentz factors $\gamma_p = 1/\sqrt{1-v_p^2}$ and $\gamma_e = 1/\sqrt{1-v_e^2}$. To retain accuracy we document $\gamma_p - 1$ and $\gamma_e - 1$.

Table IV gives all the parameters for setting up the test problems. Problem 1 in Table IV corresponds to a circularly polarized Alfven wave and is used for the accuracy analysis presented in this Sub-section. Problem 2 in Table IV corresponds to a circularly polarized electromagnetic wave and is used for the accuracy analysis presented in the next Sub-section. The problems are run to a stopping time of $t_{stop}$ which is also documented in Table IV. The Alfven wave in Problem 1 propagates with a phase speed of 0.5770370892536981. This is a strongly relativistic speed, even if it is somewhat sub-luminal (~58% of the speed of light). The electromagnetic wave in Problem 2 propagates with a phase speed of 6.250658616577547, which is a strongly super-luminal phase speed. We see, therefore, that both the test problems that we use in our accuracy analysis are strongly relativistic.

**Table IV with values for specifying circularly polarized Alfven waves or circularly polarized electromagnetic waves.**

|  | Problem 1 | Problem 2 |
|---|---|---|
| $\lambda_x$ | 64π | 16 π |
| $q_p/m_p$ | 2.876813695875796E-1 | 2.876813695875796E-1 |
| $q_e/m_e$ | -2.876813695875796E-1 | -2.876813695875796E-1 |
| $\gamma_p - 1$ | 1.538304277293179E-7 | 2.588624321120392E-6 |
| $\gamma_e - 1$ | 1.807699463451939E-7 | 3.499464816014708E-5 |
| $B_0$ | 3.615110709083961 | 3.615110709083961 |
| $\tilde{B}$ | 3.615110709083961E-3 | 3.615110709083961E-3 |
| $\tilde{E}$ | 2.086052960899683E-3 | 2.259682290361743E-2 |
| $\omega$ | 4.032169240968386E-2 | 1.747112196351524 |
| $t_{stop}$ | 312.0 | 18.0 |

When specifying the magnetic and electric fields it helps to use a vector potential approach so that we have $\mathbf{B}(x,y,z,t) = -\nabla \times \mathbf{A}(x,y,z,t)$ and $\mathbf{E}(x,y,z,t) = -\nabla \times \mathbf{C}(x,y,z,t)$. The vector potentials are explicitly given by

$$\mathbf{A}(x,y,z,t) = -B_0 z \sin\theta \hat{x} + B_0 z \cos\theta \hat{y} + \left(\tilde{B}/k^2\right)\left(k_y \hat{x} - k_x \hat{y}\right)\cos\left(k_x x + k_y y - \omega t\right) + \left(\tilde{B}/k\right)\hat{z}\sin\left(k_x x + k_y y - \omega t\right)$$

and

$$\mathbf{C}(x,y,z,t) = \left(\tilde{E}/k^2\right)\left(k_y \hat{x} - k_x \hat{y}\right)\sin\left(k_x x + k_y y - \omega t\right) - \left(\tilde{E}/k\right)\hat{z}\cos\left(k_x x + k_y y - \omega t\right)$$

With the help of these vector potentials, a divergence-free electric and magnetic field can always be set up on the computational mesh.

Table V shows the accuracy analysis for the second order scheme for Problem 1 from Table IV. We used the piecewise linear parts from a centered r=3 WENO reconstruction along with an IMEX-SSP2(3,2,2) time-integration. The errors and accuracy in the y-component of the electric and magnetic fields are shown at the last time point in the simulation; the stopping time is catalogued in Table IV. Table V shows that the method achieves second order accuracy Table VI shows the accuracy analysis for the third order scheme for the same test problem. We used the full centered r=3 WENO reconstruction along with an IMEX-SSP3(4,3,3) time-integration. Table VI shows that the method achieves third order accuracy. Table VII shows the accuracy analysis for the spatially fourth order, but temporally third order scheme for the same test problem. We used the full centered r=4 WENO reconstruction along with an IMEX-SSP3(4,3,3) time-integration. Table VII shows that the overall scheme is only third order accurate. However, notice that the value of the accuracy shown in Table VII is much better than the accuracy shown in Table VI. This shows that there is an overall benefit to using a fourth order accurate spatial reconstruction, even if the time accuracy is restricted to third order. The simulation in Table V used a CFL of 0.15 while the simulations reported in Tables VI and VII used a CFL of 0.2.

**Table V shows the accuracy analysis for the second order scheme for the propagation of an Alfven wave in a relativistic two-fluid plasma. A CFL of 0.15 was used. The errors and accuracy in the y-component of the electric and magnetic fields are shown.**

| zones | Ey $L_1$ error | Ey $L_1$ accuracy | Ey $L_{inf}$ error | Ey $L_{inf}$ accuracy |
|---|---|---|---|---|
| 32x16 | 1.1472E-4 | | 1.8059E-4 | |
| 64x32 | 2.5157E-5 | 2.19 | 3.9499E-5 | 2.19 |
| 128x64 | 5.9792E-6 | 2.07 | 9.3890E-6 | 2.07 |
| 256x128 | 1.4708E-6 | 2.02 | 2.3102E-6 | 2.02 |
| zones | By $L_1$ error | By $L_1$ accuracy | By $L_{inf}$ error | By $L_{inf}$ accuracy |
| 32x16 | 2.1652E-4 | | 3.3970E-4 | |
| 64x32 | 5.0706E-5 | 2.09 | 7.9689E-5 | 2.09 |
| 128x64 | 1.2294E-5 | 2.04 | 1.9316E-5 | 2.04 |
| 256x128 | 3.0469E-6 | 2.01 | 4.7862E-6 | 2.01 |

**Table VI shows the accuracy analysis for the third order scheme for the propagation of an Alfven wave in a relativistic two-fluid plasma. A CFL of 0.2 was used. The errors and accuracy in the y-component of the electric and magnetic fields are shown.**

| zones | Ey $L_1$ error | Ey $L_1$ accuracy | Ey $L_{inf}$ error | Ey $L_{inf}$ accuracy |
|---|---|---|---|---|

| zones | | | | |
|---|---|---|---|---|
| 32x16 | 5.4966E-5 | | 8.5912E-5 | |
| 64x32 | 7.1631E-6 | 2.94 | 1.1199E-5 | 2.94 |
| 128x64 | 8.9731E-7 | 3.00 | 1.4085E-6 | 2.99 |
| 256x128 | 1.1217E-7 | 3.00 | 1.7627E-7 | 3.00 |
| zones | By $L_1$ error | By $L_1$ accuracy | By $L_{inf}$ error | By $L_{inf}$ accuracy |
| 32x16 | 9.7564E-5 | | 1.5159E-4 | |
| 64x32 | 1.2684E-5 | 2.94 | 1.9814E-5 | 2.94 |
| 128x64 | 1.5868E-6 | 3.00 | 2.4884E-6 | 2.99 |
| 256x128 | 1.9813E-7 | 3.00 | 3.1131E-7 | 3.00 |

**Table VII shows the accuracy analysis for the spatially fourth order scheme for the propagation of an Alfven wave in a relativistic two-fluid plasma. The time accuracy was third order. A CFL of 0.2 was used. The errors and accuracy in the y-component of the electric and magnetic fields are shown.**

| zones | Ey $L_1$ error | Ey $L_1$ accuracy | Ey $L_{inf}$ error | Ey $L_{inf}$ accuracy |
|---|---|---|---|---|
| 32x16 | 3.5139E-6 | | 6.8650E-6 | |
| 64x32 | 2.2023E-7 | 4.00 | 3.8795E-7 | 4.15 |
| 128x64 | 2.1260E-8 | 3.37 | 3.4219E-8 | 3.5 |
| 256x128 | 2.4525E-9 | 3.12 | 3.8610E-9 | 3.15 |
| zones | By $L_1$ error | By $L_1$ accuracy | By $L_{inf}$ error | By $L_{inf}$ accuracy |
| 32x16 | 4.1909E-6 | | 8.9436E-6 | |
| 64x32 | 1.3470E-7 | 4.96 | 2.6752E-7 | 5.06 |
| 128x64 | 5.6274E-9 | 4.58 | 8.6452E-9 | 4.95 |
| 256x128 | 2.9115E-10 | 4.27 | 4.6079E-10 | 4.23 |

**VII.c) Two Dimensional Circularly Polarized Electromagnetic Wave in Relativistic Two-Fluid Plasma**

In this sub-section, we use a circularly polarized electromagnetic wave that propagates through a relativistic two-fluid plasma. This is documented as Problem 2 in Table IV. Table VIII shows the accuracy analysis for the second order scheme for Problem 2 from Table IV. We used the piecewise linear parts from a centered r=3 WENO reconstruction along with an IMEX-SSP2(3,2,2) time-integration. The errors and accuracy in the y-component of the electric and magnetic fields are shown at the last time point in the simulation. Table VIII shows that the method achieves second order accuracy Table IX shows the accuracy analysis for the third order scheme for the same test problem. We used the full centered r=3 WENO reconstruction along with an IMEX-SSP3(4,3,3) time-integration. Table IX shows that the method achieves third order accuracy. Table X shows the accuracy analysis for the fourth order scheme for the same test problem. We used the full centered r=4 WENO reconstruction along with an IMEX-SSP3(4,3,3) time-integration. Table X shows that the overall scheme is only third order accurate. However, notice that the value of the accuracy shown in Table X is much better than the accuracy shown in Table IX. This shows that there is an overall benefit to using a fourth order accurate spatial reconstruction, even if the time accuracy is restricted to third order. All the simulations reported in Tables VIII, IX and X used a robust CFL of 0.45.

**Table VIII shows the accuracy analysis for the second order scheme for the propagation of an electromagnetic wave in a relativistic two-fluid plasma. A CFL of 0.45 was used. The errors and accuracy in the y-component of the electric and magnetic fields are shown.**

| zones | $E_y$ $L_1$ error | $E_y$ $L_1$ accuracy | $E_y$ $L_{inf}$ error | $E_y$ $L_{inf}$ accuracy |
|---|---|---|---|---|
| 32x16 | 1.1981E-2 | | 1.8722E-2 | |
| 64x32 | 3.4557E-3 | 1.79 | 5.4265E-3 | 1.79 |
| 128x64 | 8.8987E-4 | 1.96 | 1.3979E-3 | 1.96 |
| 256x128 | 2.2166E-4 | 2.01 | 3.4818E-4 | 2.01 |
| zones | $B_y$ $L_1$ error | $B_y$ $L_1$ accuracy | $B_y$ $L_{inf}$ error | $B_y$ $L_{inf}$ accuracy |
| 32x16 | 2.0116E-3 | | 3.1628E-3 | |
| 64x32 | 5.1880E-4 | 1.96 | 8.1409E-4 | 1.96 |
| 128x64 | 1.2678E-4 | 2.03 | 1.9918E-4 | 2.03 |
| 256x128 | 3.2317E-5 | 1.97 | 5.0765E-5 | 1.97 |

**Table IX shows the accuracy analysis for the third order scheme for the propagation of an electromagnetic wave in a relativistic two-fluid plasma. A CFL of 0.45 was used. The errors and accuracy in the y-component of the electric and magnetic fields are shown.**

| zones | Ey $L_1$ error | Ey $L_1$ accuracy | Ey $L_{inf}$ error | Ey $L_{inf}$ accuracy |
|---|---|---|---|---|
| 32x16 | 1.3529E-3 | | 2.1477E-3 | |
| 64x32 | 1.3418E-4 | 3.33 | 2.2031E-4 | 3.29 |
| 128x64 | 1.5104E-5 | 3.15 | 2.4640E-5 | 3.16 |
| 256x128 | 1.8482E-6 | 3.03 | 3.0790E-6 | 3.00 |
| zones | By $L_1$ error | By $L_1$ accuracy | By $L_{inf}$ error | By $L_{inf}$ accuracy |
| 32x16 | 2.2622E-4 | | 3.5467E-4 | |
| 64x32 | 2.2140E-5 | 3.35 | 3.5224E-5 | 3.33 |
| 128x64 | 2.4266E-6 | 3.19 | 3.8805E-6 | 3.18 |
| 256x128 | 2.9141E-7 | 3.06 | 4.6825E-7 | 3.05 |

**Table X shows the accuracy analysis for the spatially fourth order scheme for the propagation of an electromagnetic wave in a relativistic two-fluid plasma. The time accuracy was third order. A CFL of 0.45 was used. The errors and accuracy in the y-component of the electric and magnetic fields are shown.**

| zones | Ey $L_1$ error | Ey $L_1$ accuracy | Ey $L_{inf}$ error | Ey $L_{inf}$ accuracy |
|---|---|---|---|---|
| 32x16 | 1.2270E-3 | | 1.9466E-3 | |
| 64x32 | 5.4427E-5 | 4.49 | 9.0256E-5 | 4.43 |
| 128x64 | 3.1089E-6 | 4.13 | 5.9289E-6 | 3.93 |
| 256x128 | 1.9290E-7 | 4.01 | 1.3098E-6 | 2.18 |
| zones | By $L_1$ error | By $L_1$ accuracy | By $L_{inf}$ error | By $L_{inf}$ accuracy |
| 32x16 | 2.0710E-4 | | 3.2745E-4 | |
| 64x32 | 9.1141E-6 | 4.51 | 1.4537E-5 | 4.49 |
| 128x64 | 4.9566E-7 | 4.20 | 8.4502E-7 | 4.10 |
| 256x128 | 3.0446E-8 | 4.03 | 5.9934E-8 | 3.82 |

## VIII) Test Problems

We present four stringent test problems here. Except for the last test problem, the resistivity was taken to be zero. The skin depth, $d_e$, is the length over which dispersive plasma effects show up and this length scale is very useful for appreciating the test problems that are presented. We define the skin depth as $d_e = c/(\Omega/\sqrt{2})$ where the plasma frequency $\Omega$ is defined after eqn. (2.23).

## VIII.a) Relativistic Brio-Wu Test Problem with Finite Plasma Skin Depth

The Brio-Wu shock tube test problem derives from classical MHD (Brio and Wu [28]). In Balsara [19] we extended this test problem to RMHD. Here we extend it further to relativistic two-fluid MHD. The problem is set up on a unit one-dimensional domain $x \in [-0.5, 0.5]$. Initially, the density of the positrons and electrons is set to 0.5 for $x < 0$ and 0.0625 for $x > 0$. The pressure of the positrons and electrons is 0.5 for $x < 0$ and 0.05 for $x > 0$. In other words, the densities and pressures on either side of the shock tube add up to the total density and pressure from the traditional RMHD version of the Brio-Wu problem. The initial velocities are all zero, as is the initial electric field. The x-component of the magnetic field is constant and set to $\sqrt{\pi}$. The y-component of the magnetic field is set to $\sqrt{4\pi}$ for $x < 0$ and $-\sqrt{4\pi}$ for $x > 0$. The z-component of the magnetic field is set to zero. We set $q_p/m_p = -q_e/m_e = 10^3/\sqrt{4\pi}$ which results in a plasma skin depth of $10^{-3}/\sqrt{\rho_p}$ where $\rho_p$ is the density of the positrons, which usually equals the electron density. As a result, when the zone size is larger than the plasma skin depth we expect to retrieve the RMHD result. For example, on a one-dimensional mesh with 400 zones, we expect to retrieve the RMHD result. However, when the zone size is much smaller than the skin depth, we expect to see oscillations associated with the dispersive waves that arise in a two-fluid plasma. As a further example, when this problem is run on a one-dimensional mesh with 1,600 zones, we expect to see dispersive waves. For this reason, we run this problem on one-dimensional meshes with 400, 1,600 and 6,400 zones. The problem was run to a final time of 0.4 with a CFL of 0.8 on all the meshes.

Fig. 3 shows the total density from the relativistic two-fluid Brio-Wu test problem. The black curve corresponds to results from a 400 zone mesh. The blue curve corresponds to results from a 1,600 zone mesh. The red curve corresponds to results from a 6,400 zone mesh. A second order scheme with MC$_\beta$ limiter was used with $\beta = 1.5$. We see that that results from the 400 zone mesh, i.e. the black curve, look exactly like those from single-fluid RMHD. The wiggles show up in the blue curve, i.e. for the 1,600 zone simulation. The red curve, i.e. the results from the 6,400 zone simulation, match quite well to the results from the blue curve. This shows that the method successfully captures two-fluid plasma effects when the resolution permits it.

**VIII.b) Relativistic Two-fluid Blast Problem with Finite Plasma Skin Depth**

The non-relativistic version of this problem was presented in Balsara & Spicer [5] and the RMHD version was presented in Komissarov [50]. We adapt this problem for relativistic two-fluid electrodynamics. The problem consists of setting up a uniform mesh on a two-dimensional computational domain spanning $[-6,6] \times [-6,6]$ and having continuitive boundary conditions. We used a mesh with $500 \times 500$ zones. Within a radius of 0.8, both the positrons and electrons have a density of $5 \times 10^{-3}$ and a pressure of $0.5$. Outside a radius of 1, both positrons and electrons have a density of $5 \times 10^{-5}$ and a pressure of $2.5 \times 10^{-4}$. A linear taper is applied to the density and pressure from a radius of 0.8 to 1, with the result that both the density and pressure linearly decrease with increasing radius in that range of radii. The initial electric field is zero and the initial magnetic field is initialized in the x-direction and has a magnitude of $0.1\sqrt{4\pi}$. We set $q_p/m_p = -q_e/m_e = 10^3/\sqrt{4\pi}$ which gives an initial skin depth of $0.1\sqrt{2}$ for radii greater than unity and a skin depth of $0.01\sqrt{2}$ for radii less than 0.8. This should be compared to a zone size of 0.024. The problem is run to a final time of 4 units with a CFL of 0.45.

Figs. 4a and 4b show the log to the base 10 of the total density and total gas pressure. Figs. 4c and 4d show the Lorentz factor for the positrons (the electrons have a comparable Lorentz factor) and the magnitude of the magnetic field. The third order accurate scheme was used. We see that the MHD result is properly reproduced. This is expected because our mesh had a zone size that is comparable to the skin depth.

It is interesting to ask whether all the ingredients of the current algorithm are necessary. In other words, it is possible to write eqn. (4.1) in conservation law form. It is, therefore, possible to treat two-fluid electrodynamics as a conservation law with sixteen zone-centered variables. We can then design a plain vanilla second order Godunov scheme with six zone-centered electric and magnetic field components, in addition to the ten zone-centered fluid variables for the positively and negatively charged fluids. We constructed such a second order accurate Godunov code with sixteen zone-centered variables and used it to rerun this problem. The results are shown in Fig. 5. Figs. 5a and 5b show the log to the base 10 of the total density and total gas pressure. Fig. 5c shows the absolute value of the undivided divergence of the magnetic field. Fig. 5d shows a zoomed view of the sub-region of Fig. 5c that spans $[-2,2]\times[3,5]$. We see that there are four locations in the density and pressure variables where there are very prominent discrepancies. Other smaller wiggles can also be seen in the region immediately behind the outermost fast magnetosonic shock. Figs. 5c and 5d show that these locations correlate quite well with the locations where the divergence of the magnetic field becomes large. Not every location with a large divergence in the magnetic field might show a problem in the flow variables. However, the locations where we observe problems in the flow do indeed correlate very well with locations where the divergence becomes large. Observe too that the undivided divergence has only to become as large as ~0.2% to 2% of the magnetic field strength before it can exert a deleterious influence on the flow. Our observation is concordant with the finding in Balsara and Kim [23] and Mocz *et al*. [63]. This proves that the algorithm presented in this paper plays an important and beneficial role in relativistic two-fluid electrodynamics.

**VIII.c) Relativistic Orzag-Tang Test Problem**

This problem was first proposed by Orzag and Tang [65] and in its relativistic form by Dumbser and Zanotti [36] or Beckwith and Stone [26]. The problem is initialized on a uniform, periodic, two-dimensional Cartesian mesh with domain $[0,1]\times[0,1]$ and $200\times 200$ zones. Both the positrons and electrons are initialized with a uniform density of $25/(72\pi)$ and a uniform pressure of $5/(24\pi)$. We set $\Gamma = 5/3$. The initial magnetic field is given by

$$\mathbf{B}(x,y) = -\sin(2\pi y)\hat{x} + \sin(4\pi x)\hat{y}$$

The initial velocities for the positrons and electrons is given by

$$\mathbf{v}(x,y) = -\frac{1}{2}\sin(2\pi y)\hat{x} + \frac{1}{2}\sin(2\pi x)\hat{y}$$

The initial electric field is given by $-\mathbf{v}(x,y) \times \mathbf{B}(x,y)$. We set $q_p/m_p = -q_e/m_e = 10^3/\sqrt{4\pi}$ which gives an initial skin depth of $3.0 \times 10^{-3}$. The problem was run to a time of unity with a CFL of 0.45.

Figs. 6a and 6b show the total density and total gas pressure. Figs. 6c and 6d show the Lorentz factor for the positrons (the electrons have a comparable Lorentz factor) and the magnitude of the magnetic field. The spatially fourth order accurate scheme was used. We see that the MHD result is properly reproduced. This is expected because our mesh had a zone size that is comparable to the skin depth.

### VIII.d) Self-similar Current Sheet with Finite Resistivity

This problem was first proposed by Komissarov [52]. Let us assume that magnetic field has only non-zero transverse component ($\mathbf{B} = B_y \hat{y}$) and the magnetic pressure is much smaller than the constant gas pressure ($B_y^2/8\pi \ll P_p + P_e$). When the magnetic field changes its sign in relatively thin current layer, its evolution follows the diffusion equation:

$$\frac{\partial B_y}{\partial t} - \frac{\eta}{4\pi}\frac{\partial^2 B_y}{\partial x^2} = 0 \ .$$

One particular self-similar solution suggested by Komissarov [52] is given as follows

$$B_y(x,t) = B_0 \mathrm{erf}\left(\sqrt{\frac{\pi}{\eta}}\frac{x}{\sqrt{t}}\right),$$

where erf is error function. The initial y-magnetic field is given by the above equation while setting t=1.0, $B_0 = \sqrt{4\pi}$ and $\eta = 0.01 \times 4\pi$. The x- and z-components of magnetic field are zero. All the components of electric field are also zero initially ($\mathbf{E} = 0$). Each of the positron and electron fluids has an initial uniform density of 0.5 and a pressure of 25.0. Hence gas pressure is 50 times larger than the magnetic pressure. The initial fluid velocities of the positron and electron fluids are zero. The problem was set up on a 200 zone mesh with [-1.5, 1.5] domain. We set $q_p/m_p = -q_e/m_e = 10^3/\sqrt{4\pi}$ which gives an initial skin depth of $\sqrt{2} \times 10^{-3}$. It was run with a CFL of 0.8 to a final time of 9.0.

The problem was run with a fourth order scheme. Fig 7 shows y-magnetic field at initial (dashed) and final time for the current sheet problem with finite resistivity. The numerical solution at the final time (dot) exactly matches with analytic solution (solid line) at the final time. We see, therefore, that our method successfully handles resistive simulations.

**VIII.e) Relativistic Two-Fluid Version of the GEM Challenge Problem**

The two-dimensional non-relativistic GEM challenge problem was documented in Birn *et al.* [27] and Hakim, Loverich and Shumlak [42]. Amano [3] has extended it to relativistic two-fluid electrodynamics. We briefly describe the set-up and present results for this very challenging test problem. The results were obtained with the fourth order accurate algorithm described in this paper.

The problem was set up for an electron-proton plasma with $m_p/m_e = 25$ and $m_p = 1$. The problem was initialized in the *xy*-plane with extent $[-L, L] \times [-L/2, L/2]$ with $L = 12.8$ units. We use a uniform mesh with $512 \times 256$ zones. The unperturbed *y*- and *z*-components of the magnetic field were set to zero. The unperturbed *x*-component of the magnetic field was given by

$$B_x(y) = B_0 \tanh(y/d)$$

with $B_0 = \sqrt{4\pi}$ and the thickness of the current sheet given by $d = 1.0$. The initial density of the protons was given by

$$\rho_p(y) = \text{sech}^2(y/d) + 0.2$$

Because the proton to electron mass ratio is fixed, the density of the electrons is also specified. The unperturbed pressures in the electron and proton fluids were set equal and are described by

$$P_p(y) = P_e(y) = \frac{B_0^2}{16\pi}\text{sech}^2(y/d) + 0.05$$

The unperturbed *x*- and *y*-velocities in the electron and proton fluids were set to zero. The unperturbed *z*-components of the four-velocities in the electron and proton fluids were set to be such as to sustain the necessary current density for the unperturbed magnetic field; the z-components of the four-velocities in the two fluids were given by

$$\gamma_p v_{z,p} = -\gamma_e v_{z,e} = -\frac{c}{8\pi ed}\frac{B_0 \text{sech}^2(y/d)}{\rho_p(y)}$$

We set the speed of light $c = 1$ and the positron charge $e = 1/\sqrt{4\pi}$. The initial electric field was set to zero. A resistivity with $\eta = 0.01 \times 4\pi$ was used in our simulations. The ratio of specific heats was set using $\Gamma = 4/3$ in the electron and proton fluids.

Now that the unperturbed flow has been described, we describe the perturbation. The only perturbation that is introduced is in the *x*- and *y*-components of the magnetic field. The z-component of the magnetic vector potential that corresponds to this perturbation in the magnetic field is given by

$$A_z(x,y) = \alpha B_0 \cos\left(\frac{\pi x}{L}\right)\cos\left(\frac{\pi y}{L}\right)$$

We use $\alpha = 0.1$. This completes our description of the set-up.

All the flow variables were treated as periodic in the x-direction. At the top and bottom boundaries of our computational domain, we require the gradients of all electric and magnetic field components to be zero. The y-velocity was made anti-symmetric at the top and bottom y-boundaries. All other flow variables were set so as to produce a zero gradient at the top and bottom y-boundaries.

The problem was run with a CFL of 0.45. This gives us a timestep that can be as much as 45 times larger than the time-explicit timestep used by Amano [3]. We have the ability to use a substantially larger timestep because of our IMEX formulation which is implicit in the treatment of the stiff source terms that arise in this problem. Figs. 8 and 9 show the solution at times of 40 and 80 units respectively. Fig. 8a shows the total mass density, Fig. 8b shows the z-component of the magnetic field. Figs. 8c and 8d show the x-components of four-velocity for the protons and electrons respectively. White contours show the overlaid magnetic field lines. Fig. 8 corresponds to a time of 40 units. The panels in Fig. 9 show the same flow variables as Fig. 8, but at a time of 80 units. Notice that the four-velocity of the electrons has become strongly relativistic in Fig. 9. In the electron fluid, the Alfven speed can be as large as 57% of the speed of light, showing that the problem is truly relativistic. We see that the proton and electron x-velocities are oppositely-oriented, as expected. The current sheet shows considerable evolution from Fig. 8 to Fig. 9, showing that substantial amount of reconnection has occurred. We can clearly see the quadrupolar z-component of the magnetic field developing at the X-point generated by the Hall effect.

## IX) Conclusions

This paper has focused on relativistic two-fluid electrodynamics. The problem consists of coupling a positively-charged relativistic fluid to a negatively-charged relativistic fluid. Both fluids, in turn, couple to the full set of Maxwell's equations. All the coupling terms are stiff and extremely non-linear. Higher order schemes are presented with spatial accuracies of two, three and four and temporal accuracies of two and three. The spatial reconstruction was based on WENO methods; the temporal evolution was based on IMEX Runge-Kutta methods. The methods meet their design accuracies. Because of the IMEX scheme, the algorithims presented here run stably at robust CFL numbers. Three prominent innovations are reported here. The first two innovations relate to the Maxwell solver and should be relevant to all codes that solve Maxwell's equations. The third innovation pertains to an efficient recasting of IMEX schemes.

The primary variables in our Maxwell equation solver are taken to be the facially collocated components of the electric and magnetic fields. Our first innovation consists of reconstructing the magnetic field within a zone in divergence-free fashion; the electric field is reconstructed in a way that is entirely consistent with Gauss' law. We realize that the generalized Ampere law and Faraday's law are based on a Stokes law type of update. As a result, edge-averaged magnetic and electric fields are needed for the update of the facial electric and magnetic fields respectively. These edge-centered magnetic and electric fields have to be obtained in a multidimensionally upwinded fashion. This is accomplished with the use of a multitimensional Riemann solver. The update strategy ensures that the electric field is always consistent with Gauss' law and the magnetic field is always divergence-free. For electromagnetic radiation propagating in vacuum, the electric and magnetic fields remain exactly divergence-free.

Since the source term coupling is stiff, an IMEX Runge-Kutta scheme is used for treating the source terms. In the past, it had proven difficult to efficiently couple IMEX schemes with higher order finite volume schemes for hyperbolic conservation laws. Our third innovation consists of finding an efficient way of arranging the calculation so that it can be done with high order accuracy.

Several accuracy analyses are presented showing that our method meets its design accuracy in the MHD limit as well as in the limit of electromagnetic wave propagation. Several stringent test problems are also presented. The stringent test problems include the relativistic version of the GEM problem, which shows that our algorithm can successfully adapt to challenging problems in high energy astrophysics.

**Acknowledgements**

DSB acknowledges support via NSF grants NSF-ACI-1307369, NSF-DMS-1361197 and NSF-ACI-1533850. DSB also acknowledges support via NASA grant NASA-NNX 12A088G. Several simulations were performed on a cluster at UND that is run by the Center for Research Computing. Computer support on NSF's XSEDE computing resources is also acknowledged.

**Appendix A**

Section III has already helped to establish notation and catalogue the details. For that reason, we will present the third order accurate reconstruction of the electric field in a much more compact fashion here. At the right and left x-faces of the reference element, the third order accurate reconstruction of the x-component of the electric field given by

$$E^{x\pm}(y,z) = E_0^{x\pm} + E_y^{x\pm} y + E_z^{x\pm} z + E_{yy}^{x\pm}\left(y^2 - 1/12\right) + E_{zz}^{x\pm}\left(z^2 - 1/12\right) + E_{yz}^{x\pm} y\, z \tag{A.1}$$

The previous equation also serves to define the moments of the x-component of the electric field that are defined in the x-face. The linear and quadratic profiles in the above equation can be obtained by using WENO reconstruction within the x-face of interest by using the x-components of the neighboring electric fields in the neighboring x-faces (Balsara and Shu [9], Balsara *et al.* [10], [11]). The two piecewise quadratic profiles for the y-component of the electric field at the upper and lower y-faces are given by

$$E^{y\pm}(x,z) = E_0^{y\pm} + E_x^{y\pm} x + E_z^{y\pm} z + E_{xx}^{y\pm}\left(x^2 - 1/12\right) + E_{zz}^{y\pm}\left(z^2 - 1/12\right) + E_{xz}^{y\pm} x\, z \tag{A.2}$$

Similarly, the two piecewise quadratic profiles for the z-component of the electric field at the top and bottom z-faces are given by

$$E^{z\pm}(x, y) = E_0^{z\pm} + E_x^{z\pm} x + E_y^{z\pm} y + E_{xx}^{z\pm}\left(x^2 - 1/12\right) + E_{yy}^{z\pm}\left(y^2 - 1/12\right) + E_{xy}^{z\pm} x\, y \quad \text{(A.3)}$$

All these profiles have to be matched in the faces by the reconstructed electric field within the reference element.

The charge density is reconstructed in piecewise quadratic fashion as

$$\begin{aligned}\rho_c(x, y, z) &= q_0 + q_x x + q_y y + q_z z + q_{xx}\left(x^2 - 1/12\right) + q_{yy}\left(y^2 - 1/12\right) + q_{zz}\left(z^2 - 1/12\right) \\ &\quad + q_{xy} x\, y + q_{yz} y\, z + q_{xz} x\, z\end{aligned} \quad \text{(A.4)}$$

Let the electric field within the unit cube be described by the following polynomials

$$\begin{aligned}E^x(x, y, z) &= a_0 + a_x x + a_y y + a_z z + a_{xx}\left(x^2 - 1/12\right) + a_{yy}\left(y^2 - 1/12\right) + a_{zz}\left(z^2 - 1/12\right) \\ &\quad + a_{xy} x\, y + a_{yz} y\, z + a_{xz} x\, z + a_{xxx}\left(x^3 - 3x/20\right) + a_{xxy}\left(x^2 - 1/12\right) y + a_{xxz}\left(x^2 - 1/12\right) z \\ &\quad + a_{xyy} x\left(y^2 - 1/12\right) + a_{xzz} x\left(z^2 - 1/12\right) + a_{xyz} x\, y\, z\end{aligned}$$

(A.5a)

$$\begin{aligned}E^y(x, y, z) &= b_0 + b_x x + b_y y + b_z z + b_{xx}\left(x^2 - 1/12\right) + b_{yy}\left(y^2 - 1/12\right) + b_{zz}\left(z^2 - 1/12\right) \\ &\quad + b_{xy} x\, y + b_{yz} y\, z + b_{xz} x\, z + b_{yyy}\left(y^3 - 3y/20\right) + b_{xyy} x\left(y^2 - 1/12\right) + b_{yyz}\left(y^2 - 1/12\right) z \\ &\quad + b_{xxy}\left(x^2 - 1/12\right) y + b_{yzz} y\left(z^2 - 1/12\right) + b_{xyz} x\, y\, z\end{aligned}$$

(A.5b)

$$\begin{aligned}E^z(x, y, z) &= c_0 + c_x x + c_y y + c_z z + c_{xx}\left(x^2 - 1/12\right) + c_{yy}\left(y^2 - 1/12\right) + c_{zz}\left(z^2 - 1/12\right) \\ &\quad + c_{xy} x\, y + c_{yz} y\, z + c_{xz} x\, z + c_{zzz}\left(z^3 - 3z/20\right) + c_{xzz} x\left(z^2 - 1/12\right) + c_{yzz} y\left(z^2 - 1/12\right) \\ &\quad + c_{xxz}\left(x^2 - 1/12\right) z + c_{yyz}\left(y^2 - 1/12\right) z + c_{xyz} x\, y\, z\end{aligned}$$

(A.5c)

The logic for the selection of these polynomials has been explained in Balsara [8]. By considering the quadratic variations in Gauss' law we get the following consistency relations

$$3a_{xxx} + b_{xxy} + c_{xxz} = q_{xx} \quad ; \quad a_{xyy} + 3b_{yyy} + c_{yyz} = q_{yy} \quad ; \quad a_{xzz} + b_{yzz} + 3c_{zzz} = q_{zz} \quad ;$$
$$2a_{xxy} + 2b_{xyy} + c_{xyz} = q_{xy} \quad ; \quad a_{xyz} + 2b_{yyz} + 2c_{yzz} = q_{yz} \quad ; \quad 2a_{xxz} + b_{xyz} + 2c_{xzz} = q_{xz}$$
(A.6)

By considering the linear variations in Gauss' law we get consistency relations that are identical with eqn. (3.7). By considering the constant terms in Gauss' law we get the following consistency relation

$$a_x + b_y + c_z - 3(a_{xxx} + b_{yyy} + c_{zzz})/20$$
$$- (a_{xyy} + a_{xzz} + b_{xxy} + b_{yzz} + c_{xxz} + c_{yyz})/12 = q_0 - (q_{xx} + q_{yy} + q_{zz})/12$$
(A.7)

In the next paragraph we will find the coefficients that satisfy these consistency relations.

Matching the quadratically varying part of the x-component of the electric field in the right and left x-faces gives

$$a_{yy} = (E_{yy}^{x+} + E_{yy}^{x-})/2 \quad ; \quad a_{xyy} = E_{yy}^{x+} - E_{yy}^{x-} \quad ; \quad a_{zz} = (E_{zz}^{x+} + E_{zz}^{x-})/2 \quad ; \quad a_{xzz} = E_{zz}^{x+} - E_{zz}^{x-} \quad ;$$
$$a_{yz} = (E_{yz}^{x+} + E_{yz}^{x-})/2 \quad ; \quad a_{xyz} = E_{yz}^{x+} - E_{yz}^{x-}$$
(A.8)

The same can be done for the y-component of the electric field at the upper and lower y-faces. We do not write out the explicit relations because they can be obtained from the above equation with the transcription $a \to b$ and the cyclic rotation of indices given by $(x, y, z) \to (y, z, x)$. Also note that we have to identify $b_{zx}$ with $b_{xz}$ etc. A similar matching can be done for the z-component of the electric field at the top and bottom z-faces. Again, the resultant equations can be obtained by making the transcription $a \to c$ and the cyclic rotation of indices given by $(x, y, z) \to (z, x, y)$. Satisfying the consistency relations for the quadratic terms also requires us to make a least squares minimization of the electric energy; fortunately this is trivial and yields

$$a_{xxx} = -(b_{xxy} + c_{xxz} - q_{xx})/3 \quad ; \quad b_{yyy} = -(a_{xyy} + c_{yyz} - q_{yy})/3 \quad ; \quad c_{zzz} = -(a_{xzz} + b_{yzz} - q_{zz})/3 \quad ;$$
$$a_{xxy} = b_{xyy} = -(c_{xyz} - q_{xy})/4 \quad ; \quad b_{yyz} = c_{yzz} = -(a_{xyz} - q_{yz})/4 \quad ; \quad a_{xxz} = c_{xzz} = -(b_{xyz} - q_{xz})/4$$
(A.9)

Matching the linearly varying part of the x-component of the electric field in the right and left x-faces gives

$$a_y = \left(E_y^{x+} + E_y^{x-}\right)/2 - a_{xxy}/6 \quad ; \quad a_{xy} = E_y^{x+} - E_y^{x-} \quad ;$$
$$a_z = \left(E_z^{x+} + E_z^{x-}\right)/2 - a_{xxz}/6 \quad ; \quad a_{xz} = E_z^{x+} - E_z^{x-}$$
(A.10)

As before, the above equation can be used to obtain the corresponding "b" and "c" coefficients by cyclic rotations of the subscripts and superscripts. The consistency relations from eqn. (3.12) should then be imposed. Matching the constant part of the x-component of the electric field in the right and left x-faces gives

$$a_0 = \left(E_0^{x+} + E_0^{x-}\right)/2 - a_{xx}/6 \quad ; \quad a_x = E_0^{x+} - E_0^{x-} - a_{xxx}/10$$
(A.11)

As before, the above equation can be used to obtain the corresponding "b" and "c" coefficients by cyclic rotations of the subscripts and superscripts. This completes our description of the third order accurate reconstruction of the electric field in the presence of a charge density.

**Appendix B**

Section III has already helped to establish notation. For that reason, we will present the fourth order accurate reconstruction of the electric field in a much more compact fashion here. At the right and left x-faces of the reference element, the fourth order accurate reconstruction of the x-component of the electric field is given by

$$E^{x\pm}(y,z) = E_0^{x\pm} + E_y^{x\pm} y + E_z^{x\pm} z + E_{yy}^{x\pm}\left(y^2 - 1/12\right) + E_{zz}^{x\pm}\left(z^2 - 1/12\right) + E_{yz}^{x\pm} y z$$
$$+ E_{yyy}^{x\pm}\left(y^3 - 3y/20\right) + E_{zzz}^{x\pm}\left(z^3 - 3z/20\right) + E_{yyz}^{x\pm}\left(y^2 - 1/12\right)z + E_{yzz}^{x\pm} y\left(z^2 - 1/12\right)$$
(B.1)

A WENO strategy for obtaining the moments of the above equation is described in (Balsara and Shu [9], Balsara *et al*. [10], [11]). The two piecewise cubic profiles for the y-component of the electric field at the upper and lower y-faces are given by

$$E^{y\pm}(x,z) = E_0^{y\pm} + E_x^{y\pm}x + E_z^{y\pm}z + E_{xx}^{y\pm}\left(x^2 - 1/12\right) + E_{zz}^{y\pm}\left(z^2 - 1/12\right) + E_{xz}^{y\pm}x\,z$$
$$+ E_{xxx}^{y\pm}\left(x^3 - 3x/20\right) + E_{zzz}^{y\pm}\left(z^3 - 3z/20\right) + E_{xxz}^{y\pm}\left(x^2 - 1/12\right)z + E_{xzz}^{y\pm}x\left(z^2 - 1/12\right) \quad \text{(B.2)}$$

Similarly, the two piecewise cubic profiles for the z-component of the electric field at the top and bottom z-faces are given by

$$E^{z\pm}(x,y) = E_0^{z\pm} + E_x^{z\pm}x + E_y^{z\pm}y + E_{xx}^{z\pm}\left(x^2 - 1/12\right) + E_{yy}^{z\pm}\left(y^2 - 1/12\right) + E_{xy}^{z\pm}x\,y$$
$$+ E_{xxx}^{z\pm}\left(x^3 - 3x/20\right) + E_{yyy}^{z\pm}\left(y^3 - 3y/20\right) + E_{xxy}^{z\pm}\left(x^2 - 1/12\right)y + E_{xyy}^{z\pm}x\left(y^2 - 1/12\right) \quad \text{(B.3)}$$

These three profiles have to be matched in the faces by the reconstructed electric field within the reference element.

The charge density is reconstructed in piecewise cubic fashion as

$$\rho_c(x,y,z) = q_0 + q_x x + q_y y + q_z z + q_{xx}\left(x^2 - 1/12\right) + q_{yy}\left(y^2 - 1/12\right) + q_{zz}\left(z^2 - 1/12\right)$$
$$+ q_{xy}x\,y + q_{yz}y\,z + q_{xz}x\,z + q_{xxx}\left(x^3 - 3x/20\right) + q_{yyy}\left(y^3 - 3y/20\right) + q_{zzz}\left(z^3 - 3z/20\right)$$
$$+ q_{xxy}\left(x^2 - 1/12\right)y + q_{xxz}\left(x^2 - 1/12\right)z + q_{xyy}x\left(y^2 - 1/12\right)$$
$$+ q_{yyz}\left(y^2 - 1/12\right)z + q_{xzz}x\left(z^2 - 1/12\right) + q_{yzz}y\left(z^2 - 1/12\right) + q_{xyz}x\,y\,z$$

(B.4)

Let the electric field within the unit cube be described by the following polynomials

$$E^x(x,y,z) = a_0 + a_x x + a_y y + a_z z + a_{xx}\left(x^2 - 1/12\right) + a_{yy}\left(y^2 - 1/12\right) + a_{zz}\left(z^2 - 1/12\right)$$
$$+ a_{xy}x\,y + a_{yz}y\,z + a_{xz}x\,z + a_{xxx}\left(x^3 - 3x/20\right) + a_{yyy}\left(y^3 - 3y/20\right) + a_{zzz}\left(z^3 - 3z/20\right)$$
$$+ a_{xxy}\left(x^2 - 1/12\right)y + a_{xxz}\left(x^2 - 1/12\right)z + a_{xyy}x\left(y^2 - 1/12\right) + a_{yyz}\left(y^2 - 1/12\right)z$$
$$+ a_{xzz}x\left(z^2 - 1/12\right) + a_{yzz}y\left(z^2 - 1/12\right) + a_{xyz}x\,y\,z$$
$$+ a_{xyyy}x\left(y^3 - 3y/20\right) + a_{xyyz}x\left(y^2 - 1/12\right)z + a_{xyzz}x\,y\left(z^2 - 1/12\right) + a_{xzzz}x\left(z^3 - 3z/20\right)$$
$$+ a_{xxxx}\left(x^4 - 3x^2/14 + 3/560\right) + a_{xxxy}\left(x^3 - 3x/20\right)y + a_{xxxz}\left(x^3 - 3x/20\right)z$$
$$+ a_{xxyy}\left(x^2 - 1/12\right)\left(y^2 - 1/12\right) + a_{xxzz}\left(x^2 - 1/12\right)\left(z^2 - 1/12\right) + a_{xxyz}\left(x^2 - 1/12\right)y\,z$$

(B.5a)

$$\begin{aligned}
E^y(x,y,z) = {} & b_0 + b_x\,x + b_y\,y + b_z\,z + b_{xx}\left(x^2 - 1/12\right) + b_{yy}\left(y^2 - 1/12\right) + b_{zz}\left(z^2 - 1/12\right) \\
& + b_{xy}\,x\,y + b_{yz}\,y\,z + b_{xz}\,x\,z + b_{xxx}\left(x^3 - 3x/20\right) + b_{yyy}\left(y^3 - 3y/20\right) + b_{zzz}\left(z^3 - 3z/20\right) \\
& + b_{xxy}\left(x^2 - 1/12\right)y + b_{xxz}\left(x^2 - 1/12\right)z + b_{xyy}\,x\left(y^2 - 1/12\right) + b_{yyz}\left(y^2 - 1/12\right)z \\
& + b_{xzz}\,x\left(z^2 - 1/12\right) + b_{yzz}\,y\left(z^2 - 1/12\right) + b_{xyz}\,x\,y\,z \\
& + b_{xxxy}\left(x^3 - 3x/20\right)y + b_{xxyz}\left(x^2 - 1/12\right)y\,z + b_{xyzz}\,x\,y\left(z^2 - 1/12\right) + b_{yzzz}\,y\left(z^3 - 3z/20\right) \\
& + b_{yyyy}\left(y^4 - 3y^2/14 + 3/560\right) + b_{xyyy}\,x\left(y^3 - 3y/20\right) + b_{yyyz}\left(y^3 - 3y/20\right)z \\
& + b_{xxyy}\left(x^2 - 1/12\right)\left(y^2 - 1/12\right) + b_{yyzz}\left(y^2 - 1/12\right)\left(z^2 - 1/12\right) + b_{xyyz}\,x\left(y^2 - 1/12\right)z
\end{aligned}$$

(B.5b)

$$\begin{aligned}
E^z(x,y,z) = {} & c_0 + c_x\,x + c_y\,y + c_z\,z + c_{xx}\left(x^2 - 1/12\right) + c_{yy}\left(y^2 - 1/12\right) + c_{zz}\left(z^2 - 1/12\right) \\
& + c_{xy}\,x\,y + c_{yz}\,y\,z + c_{xz}\,x\,z + c_{xxx}\left(x^3 - 3x/20\right) + c_{yyy}\left(y^3 - 3y/20\right) + c_{zzz}\left(z^3 - 3z/20\right) \\
& + c_{xxy}\left(x^2 - 1/12\right)y + c_{xxz}\left(x^2 - 1/12\right)z + c_{xyy}\,x\left(y^2 - 1/12\right) + c_{yyz}\left(y^2 - 1/12\right)z \\
& + c_{xzz}\,x\left(z^2 - 1/12\right) + c_{yzz}\,y\left(z^2 - 1/12\right) + c_{xyz}\,x\,y\,z \\
& + c_{xxxz}\left(x^3 - 3x/20\right)z + c_{xxyz}\left(x^2 - 1/12\right)y\,z + c_{xyyz}\,x\left(y^2 - 1/12\right)z + c_{yyyz}\left(y^3 - 3y/20\right)z \\
& + c_{zzzz}\left(z^4 - 3z^2/14 + 3/560\right) + c_{xzzz}\,x\left(z^3 - 3z/20\right) + c_{yzzz}\,y\left(z^3 - 3z/20\right) \\
& + c_{xxzz}\left(x^2 - 1/12\right)\left(z^2 - 1/12\right) + c_{yyzz}\left(y^2 - 1/12\right)\left(z^2 - 1/12\right) + c_{xyzz}\,x\,y\left(z^2 - 1/12\right)
\end{aligned}$$

(B.5c)

By considering the cubic variations in Gauss' law we get the following consistency relations

$$\begin{aligned}
& 4a_{xxxx} + b_{xxxy} + c_{xxxz} = q_{xxx} \;;\; a_{xyyy} + 4b_{yyyy} + c_{yyyz} = q_{yyy} \;;\; a_{xzzz} + b_{yzzz} + 4c_{zzzz} = q_{zzz} \;; \\
& 3a_{xxxy} + 2b_{xxyy} + c_{xxyz} = q_{xxy} \;;\; 3a_{xxxz} + b_{xxyz} + 2c_{xxzz} = q_{xxz} \;;\; 2a_{xxyy} + 3b_{xyyy} + c_{xyyz} = q_{xyy} \;; \\
& a_{xyyz} + 3b_{yyyz} + 2c_{yyzz} = q_{yyz} \;;\; 2a_{xxzz} + b_{xyzz} + 3c_{xzzz} = q_{xzz} \;;\; a_{xyzz} + 2b_{yyzz} + 3c_{yzzz} = q_{yzz} \;; \\
& 2a_{xxyz} + 2b_{xyyz} + 2c_{xyzz} = q_{xyz}
\end{aligned}$$

(B.6)

By considering the quadratic variations in Gauss' law we retrieve the consistency relations in eqn. (A.6). By considering the linear variations in Gauss' law we get the consistency relations

$$2a_{xx} + b_{xy} + c_{xz} - 3a_{xxxx}/7 - (a_{xxyy} + a_{xxzz})/6 - 3(b_{xxxy} + b_{xyyy} + c_{xxxz} + c_{xzzz})/20$$
$$-(b_{xyzz} + c_{xyyz})/12 = q_x - 3q_{xxx}/20 - (q_{xyy} + q_{xzz})/12 \quad;$$
$$a_{xy} + 2b_{yy} + c_{yz} - 3b_{yyyy}/7 - (b_{xxyy} + b_{yyzz})/6 - 3(a_{xxxy} + a_{xyyy} + c_{yyyz} + c_{yzzz})/20$$
$$-(a_{xyzz} + c_{xyyz})/12 = q_y - 3q_{yyy}/20 - (q_{xxy} + q_{yzz})/12 \quad; \quad \text{(B.7)}$$
$$a_{xz} + b_{yz} + 2c_{zz} - 3c_{zzzz}/7 - (c_{xxzz} + c_{yyzz})/6 - 3(a_{xxxz} + a_{xzzz} + b_{yyyz} + b_{yzzz})/20$$
$$-(a_{xyyz} + b_{xxyz})/12 = q_z - 3q_{zzz}/20 - (q_{xxz} + q_{yyz})/12$$

Considering the constant terms in Gauss' law retrieves the consistency relation in eqn. (A.7).

Matching the qubically varying part of the x-component of the electric field in the right and left x-faces gives

$$a_{yyy} = \left(E_{yyy}^{x+} + E_{yyy}^{x-}\right)/2 \quad; \quad a_{xyyy} = E_{yyy}^{x+} - E_{yyy}^{x-} \quad; \quad a_{yyz} = \left(E_{yyz}^{x+} + E_{yyz}^{x-}\right)/2 \quad; \quad a_{xyyz} = E_{yyz}^{x+} - E_{yyz}^{x-} \quad;$$
$$a_{yzz} = \left(E_{yzz}^{x+} + E_{yzz}^{x-}\right)/2 \quad; \quad a_{xyzz} = E_{yzz}^{x+} - E_{yzz}^{x-} \quad; \quad a_{zzz} = \left(E_{zzz}^{x+} + E_{zzz}^{x-}\right)/2 \quad; \quad a_{xzzz} = E_{zzz}^{x+} - E_{zzz}^{x-}$$

(B.8)

The same can be done for the y-component of the electric field at the upper and lower y-faces by making the transcription $a \to b$ and the cyclic rotation of indices given by $(x, y, z) \to (y, z, x)$. A similar matching can be done for the z-component of the electric field at the top and bottom z-faces by making the transcription $a \to c$ and the cyclic rotation of indices given by $(x, y, z) \to (z, x, y)$. Satisfying the consistency relations for the cubic terms also requires us to make a least squares minimization of the electric energy. Unfortunately this is not entirely trivial; however, it can be done and yields

$$a_{xxxx} = -(b_{xxxy} + c_{xxxz} - q_{xxx})/4 \ ; \ b_{yyyy} = -(a_{xyyy} + c_{yyyz} - q_{yyy})/4 \ ; \ c_{zzzz} = -(a_{xzzz} + b_{yzzz} - q_{zzz})/4 \ ;$$

$$b_{xxyy} = -3(c_{xxyz} - q_{xxy})/20 \ ; \ a_{xxxy} = -7(c_{xxyz} - q_{xxy})/30 \ ;$$

$$c_{xxzz} = -3(b_{xxyz} - q_{xxz})/20 \ ; \ a_{xxxz} = -7(b_{xxyz} - q_{xxz})/30 \ ;$$

$$a_{xxyy} = -3(c_{xyyz} - q_{xyy})/20 \ ; \ b_{xyyy} = -7(c_{xyyz} - q_{xyy})/30 \ ;$$

$$c_{yyzz} = -3(a_{xyyz} - q_{yyz})/20 \ ; \ b_{yyyz} = -7(a_{xyyz} - q_{yyz})/30 \ ;$$

$$a_{xxzz} = -3(b_{xyzz} - q_{xzz})/20 \ ; \ c_{xzzz} = -7(b_{xyzz} - q_{xzz})/30 \ ;$$

$$b_{yyzz} = -3(a_{xyzz} - q_{yzz})/20 \ ; \ c_{yzzz} = -7(a_{xyzz} - q_{yzz})/30 \ ; \ a_{xxyz} = b_{xyyz} = c_{xyzz} = q_{xyz}/6$$

(B.9)

Matching the quadratically varying part of the x-component of the electric field in the right and left x-faces gives

$$a_{yy} = (E_{yy}^{x+} + E_{yy}^{x-})/2 - a_{xxyy}/6 \ ; \ a_{xyy} = E_{yy}^{x+} - E_{yy}^{x-} \ ; \ a_{zz} = (E_{zz}^{x+} + E_{zz}^{x-})/2 - a_{xxzz}/6 \ ; \ a_{xzz} = E_{zz}^{x+} - E_{zz}^{x-} \ ;$$

$$a_{yz} = (E_{yz}^{x+} + E_{yz}^{x-})/2 - a_{xxyz}/6 \ ; \ a_{xyz} = E_{yz}^{x+} - E_{yz}^{x-}$$

(B.10)

As before, the above equation can be used to obtain the corresponding "b" and "c" coefficients by cyclic rotations of the subscripts and superscripts. The consistency relations from eqn. (A.9) should then be imposed. Matching the linearly varying part of the x-component of the electric field in the right and left x-faces gives

$$a_y = (E_y^{x+} + E_y^{x-})/2 - a_{xxy}/6 \ ; \quad a_{xy} = E_y^{x+} - E_y^{x-} - a_{xxxy}/10 \ ;$$

$$a_z = (E_z^{x+} + E_z^{x-})/2 - a_{xxz}/6 \ ; \quad a_{xz} = E_z^{x+} - E_z^{x-} - a_{xxxz}/10$$

(B.11)

As before, the above equation can be used to obtain the corresponding "b" and "c" coefficients by cyclic rotations of the subscripts and superscripts. The consistency relations from eqn. (B.7) then give us the additional coefficients

$$\begin{aligned}
a_{xx} &= -(b_{xy}+c_{xz})/2 + 3a_{xxxx}/14 + (a_{xyyy}+a_{xxzz})/12 + 3(b_{xxxy}+b_{xyyy}+c_{xxxz}+c_{xzzz})/40 \\
&\quad + (b_{xyzz}+c_{xyyz})/24 + q_x/2 - 3q_{xxx}/40 - (q_{xyy}+q_{xzz})/24 \\
b_{yy} &= -(a_{xy}+c_{yz})/2 + 3b_{yyyy}/14 + (b_{xxyy}+b_{yyzz})/12 + 3(a_{xxxy}+a_{xyyy}+c_{yyyz}+c_{yzzz})/40 \\
&\quad + (a_{xyzz}+c_{xxyz})/24 + q_y/2 - 3q_{yyy}/40 - (q_{xxy}+q_{yzz})/24 \\
c_{zz} &= -(a_{xz}+b_{yz})/2 + 3c_{zzzz}/14 + (c_{xxzz}+c_{yyzz})/12 + 3(a_{xxxz}+a_{xzzz}+b_{yyyz}+b_{yzzz})/40 \\
&\quad + (a_{xyyz}+b_{xxyz})/24 + q_z/2 - 3q_{zzz}/40 - (q_{xxz}+q_{yyz})/24
\end{aligned} \qquad (B.12)$$

Matching the constant part of the x-component of the electric field in the right and left x-faces gives

$$a_0 = \left(E_0^{x+} + E_0^{x-}\right)/2 - a_{xx}/6 - a_{xxxx}/70 \quad ; \quad a_x = E_0^{x+} - E_0^{x-} - a_{xxx}/10 \qquad (A.13)$$

As before, the above equation can be used to obtain the corresponding "b" and "c" coefficients by cyclic rotations of the subscripts and superscripts. This completes our description of the fourth order accurate reconstruction of the electric field in the presence of a charge density.

**Appendix C**

Here we give explicit expressions for all components of the electric and magnetic field from a multidimensional Riemann solver for electrodynamics on a Cartesian mesh. We slightly expand the notation so that a subscript of $(x+)$ denotes a variable that is taken from the positive x-direction of a vertex being considered while a subscript of $(x-)$ denotes a variable that is taken from the negative x-direction of the same vertex. Similar considerations apply to the other coordinate directions. As a result, eqns. (4.2) and (4.3) become

$$E^{z*} = \frac{1}{4}\left(E^z_{(x+)(y+)} + E^z_{(x+)(y-)} + E^z_{(x-)(y+)} + E^z_{(x-)(y-)}\right) + \frac{1}{2}\left(B^{y*}_{(x+)} - B^{y*}_{(x-)}\right) - \frac{1}{2}\left(B^{x*}_{(y+)} - B^{x*}_{(y-)}\right) \qquad (C.1)$$

$$B^{z*} = \frac{1}{4}\left(B^z_{(x+)(y+)} + B^z_{(x+)(y-)} + B^z_{(x-)(y+)} + B^z_{(x-)(y-)}\right) - \frac{1}{2}\left(E^{y*}_{(x+)} - E^{y*}_{(x-)}\right) + \frac{1}{2}\left(E^{x*}_{(y+)} - E^{x*}_{(y-)}\right) \qquad (C.2)$$

Making the cyclic transcription $(x, y, z) \to (z, x, y)$ gives

$$E^{y*} = \frac{1}{4}\left(E^{y}_{(x+)(z+)} + E^{y}_{(x+)(z-)} + E^{y}_{(x-)(z+)} + E^{y}_{(x-)(z-)}\right) + \frac{1}{2}\left(B^{x*}_{(z+)} - B^{x*}_{(z-)}\right) - \frac{1}{2}\left(B^{z*}_{(x+)} - B^{z*}_{(x-)}\right) \quad \text{(C.3)}$$

$$B^{y*} = \frac{1}{4}\left(B^{y}_{(x+)(z+)} + B^{y}_{(x+)(z-)} + B^{y}_{(x-)(z+)} + B^{y}_{(x-)(z-)}\right) - \frac{1}{2}\left(E^{x*}_{(z+)} - E^{x*}_{(z-)}\right) + \frac{1}{2}\left(E^{z*}_{(x+)} - E^{z*}_{(x-)}\right) \quad \text{(C.4)}$$

Applying the cyclic transcription $(x, y, z) \to (y, z, x)$ to eqns. (C.1) and (C.2) gives

$$E^{x*} = \frac{1}{4}\left(E^{x}_{(y+)(z+)} + E^{x}_{(y+)(z-)} + E^{x}_{(y-)(z+)} + E^{x}_{(y-)(z-)}\right) + \frac{1}{2}\left(B^{z*}_{(y+)} - B^{z*}_{(y-)}\right) - \frac{1}{2}\left(B^{y*}_{(z+)} - B^{y*}_{(z-)}\right) \quad \text{(C.5)}$$

$$B^{x*} = \frac{1}{4}\left(B^{x}_{(y+)(z+)} + B^{x}_{(y+)(z-)} + B^{x}_{(y-)(z+)} + B^{x}_{(y-)(z-)}\right) - \frac{1}{2}\left(E^{z*}_{(y+)} - E^{z*}_{(y-)}\right) + \frac{1}{2}\left(E^{y*}_{(z+)} - E^{y*}_{(z-)}\right) \quad \text{(C.6)}$$

This completes our description of the electromagnetic fields that are provided at the edges of the mesh using the multidimensional Riemann solver. The edge-centered electric fields from eqns. (C.1), (C.3) and (C.5) enable us to update face-centered magnetic fields using Faraday's law. The edge-centered magnetic fields from eqns. (C.2), (C.4) and (C.6) enable us to update the face-centered electric fields using the generalized Ampere's law.

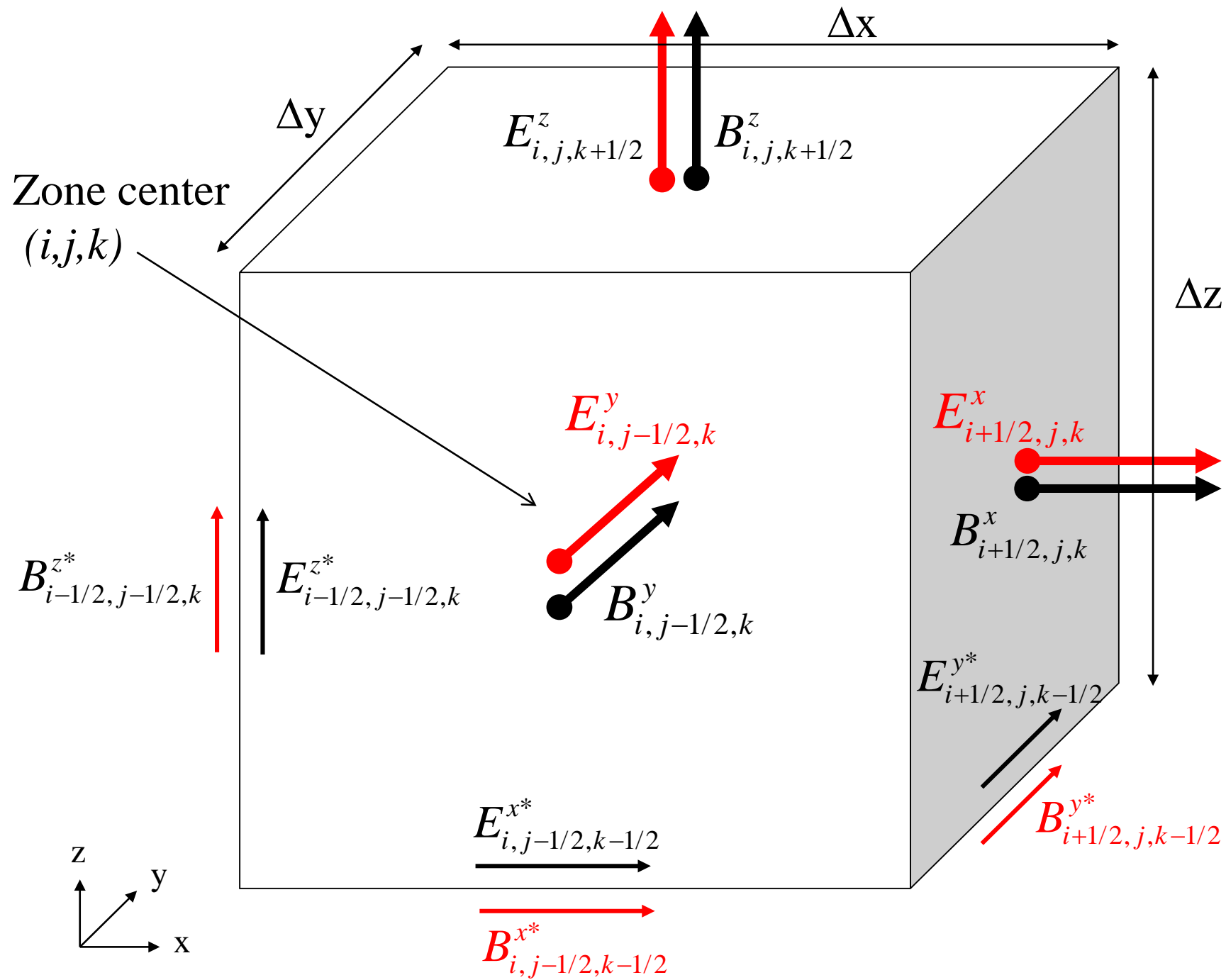

*Fig. 1 shows us that the primary magnetic and electric field variables are facially-collocated and undergo an update from Faraday's law and the generalized Ampere's law respectively. The components of the primary magnetic field are shown by the thick black arrows while the components of the primary electric field are shown by the thick red arrows. The edge-centered electric fields, which are used for updating the facial magnetic field components, are shown by the thin black arrows close to the appropriate edge. The edge-centered magnetic fields, which are used for updating the facial electric field components, are shown by the thin red arrows close to the appropriate edge.*

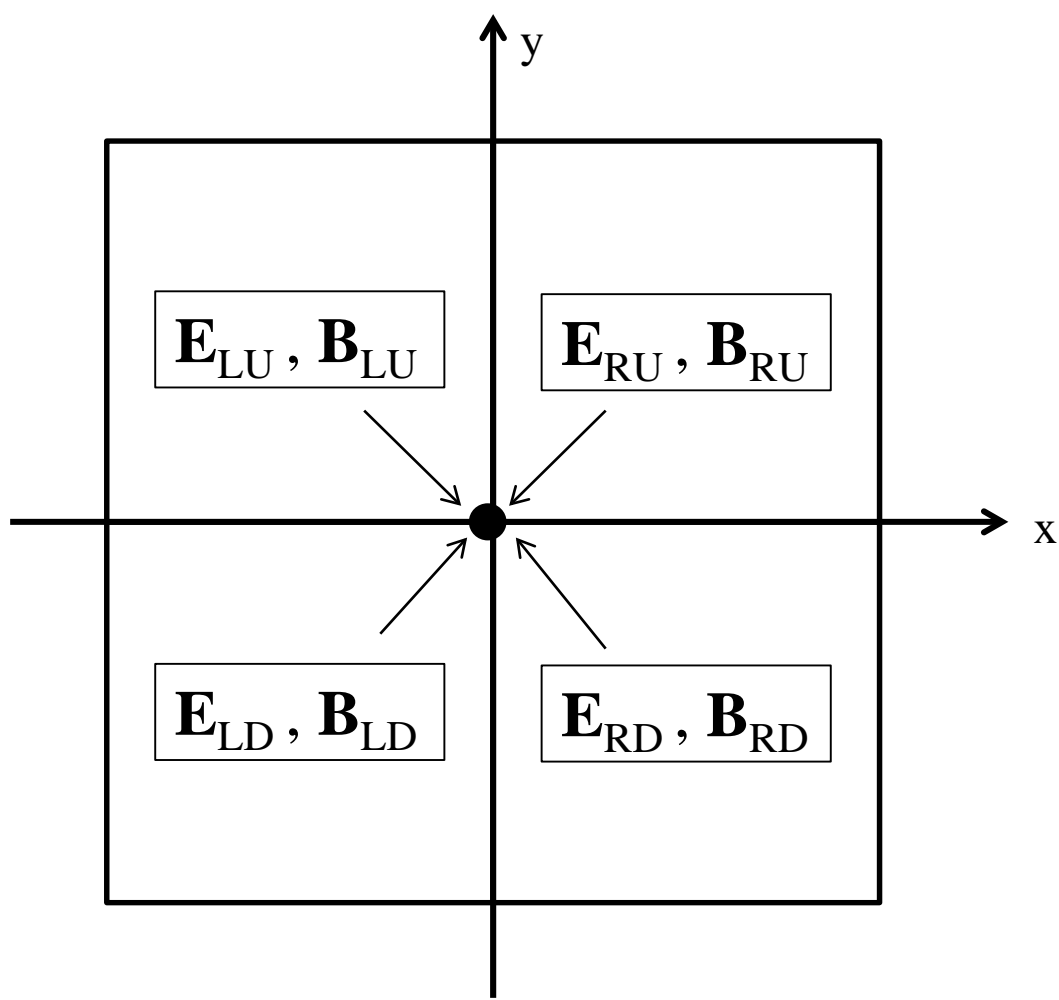

Fig. 2 shows four zones in the xy-plane that come together at the z-edge of a three-dimensional mesh. Since the mesh is viewed from the top in plan view, the z-edge is shown by the black dot and the four abutting zones are shown as four squares. The four states have subscripts given by "RU" for right-upper; "LU" for left-upper; "LD" for left-down and "RD" for right-down.

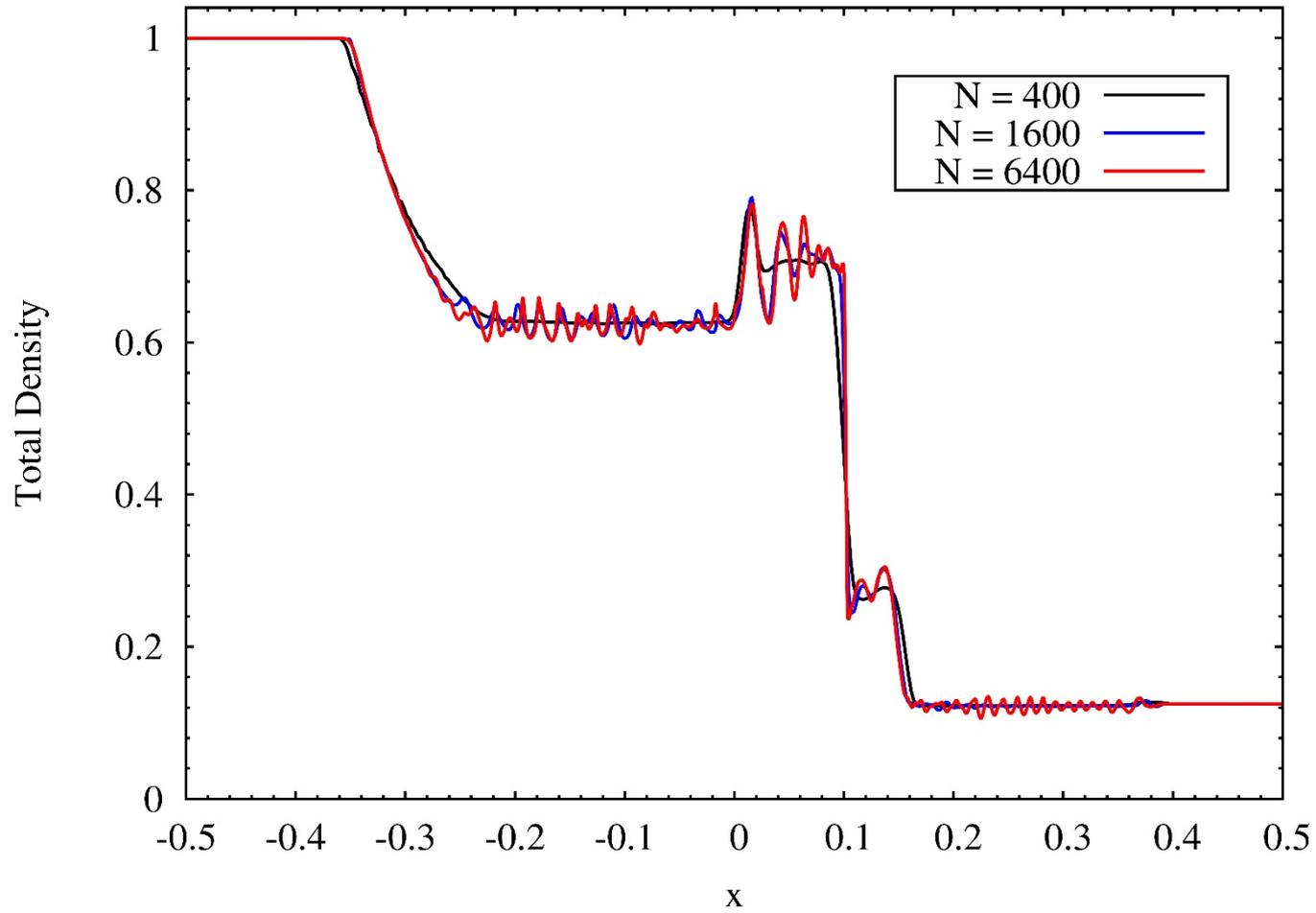

*Fig. 3 shows the total density from the relativistic two-fluid Brio-Wu test problem. The black curve corresponds to results from a 400 zone mesh. The blue curve corresponds to results from a 1,600 zone mesh. The red curve corresponds to results from a 6,400 zone mesh. A second order scheme with $MC_\beta$ limiter was used.*

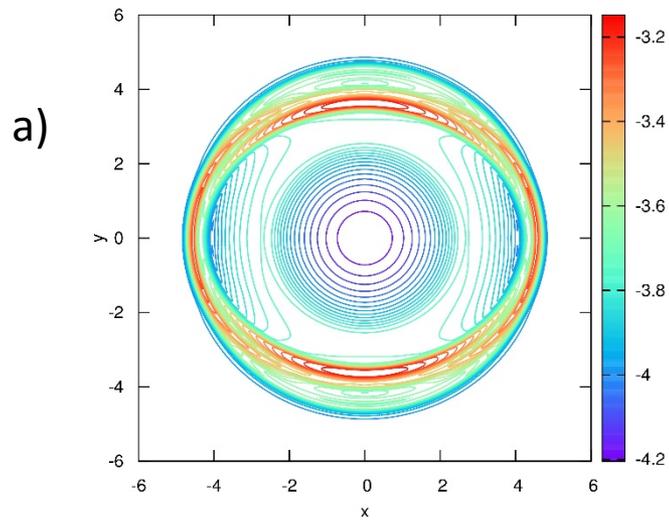 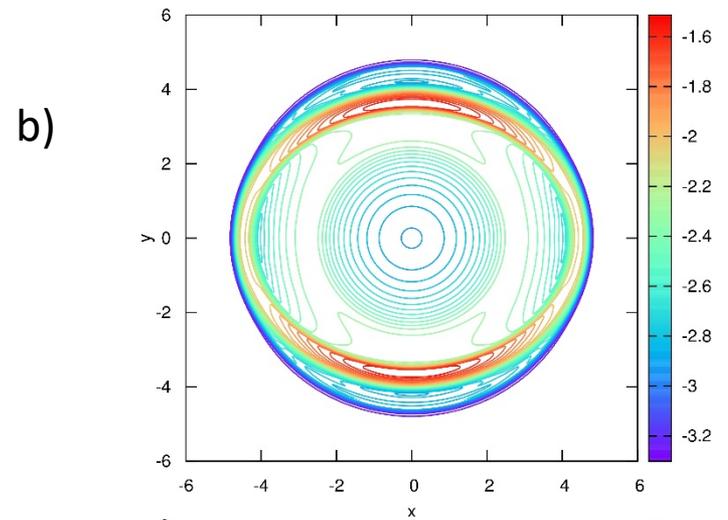
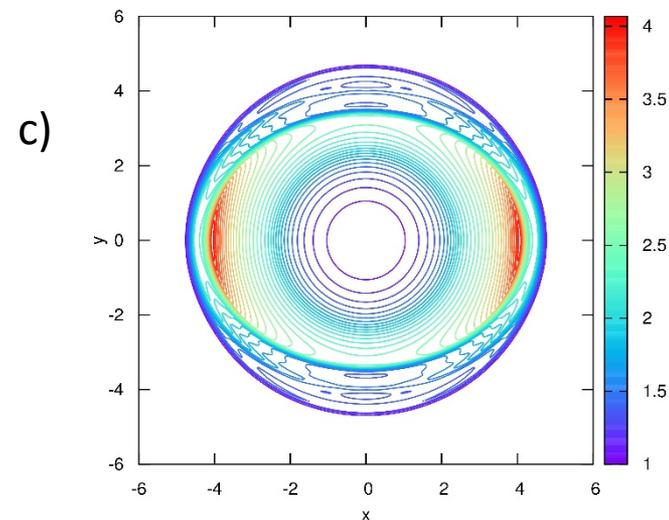 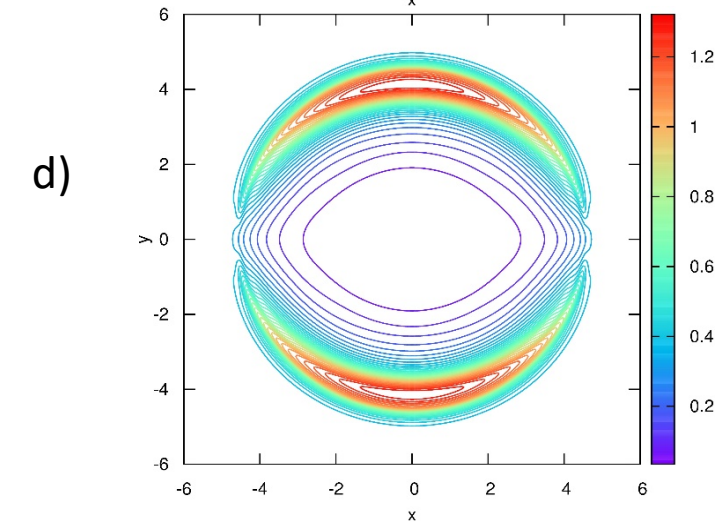

*Figs. 4a and 4b show the log$_{10}$ of the total density and total gas pressure. Figs. 4c and 4d show the Lorentz factor for the positrons (the electrons have a comparable Lorentz factor) and the magnitude of the magnetic field. A third order scheme was used.*

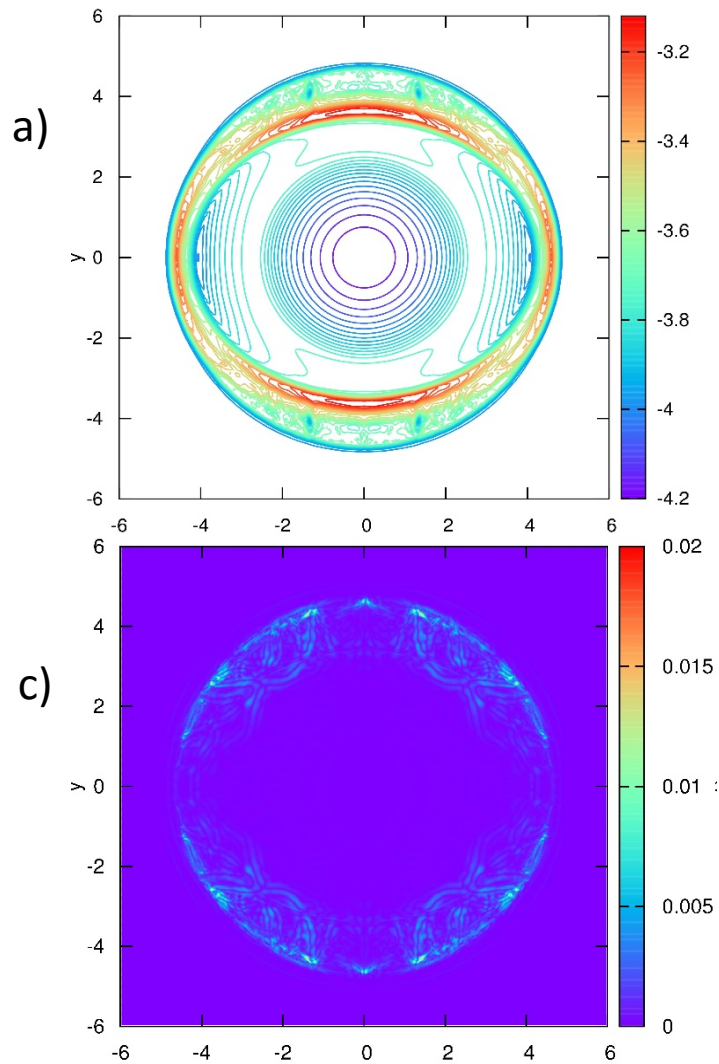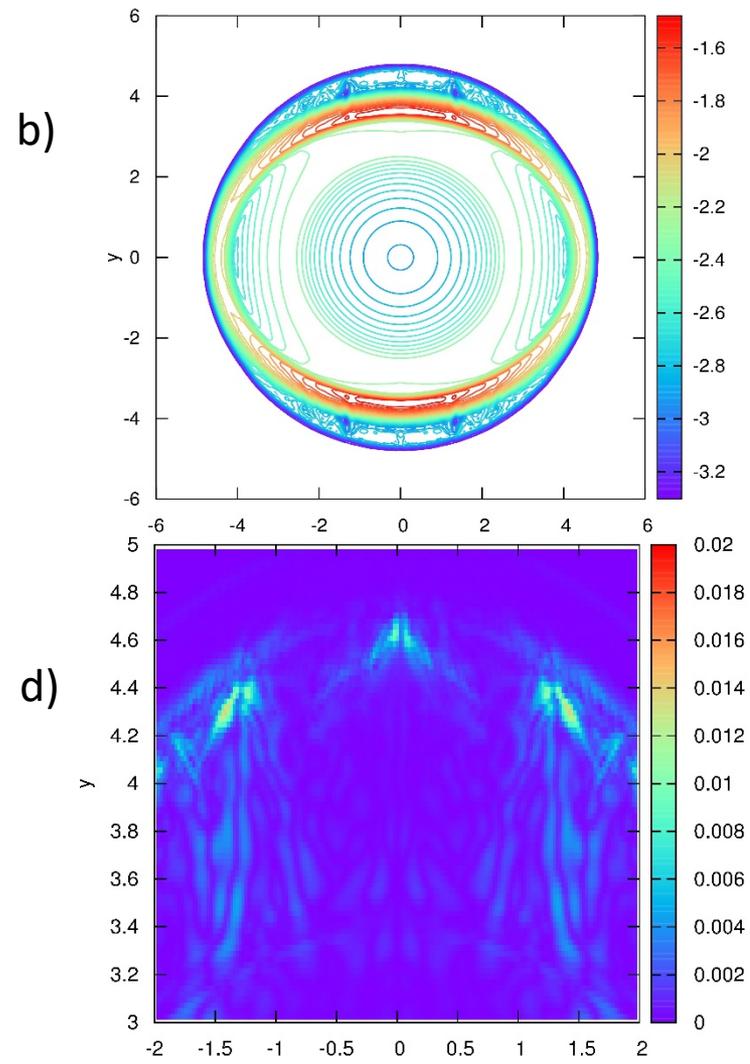

*Figs. 5a and 5b show the log$_{10}$ of the density and pressure. Fig. 5c shows the absolute value of the undivided divergence of the magnetic field. Fig. 5d zooms into a small subsection of Fig. 5c. In this figure a second order Godunov scheme with zone-centered electric and magnetic fields was used. The discrepancies in the density and pressure correlate with the locations where the divergence of the magnetic field builds up.*

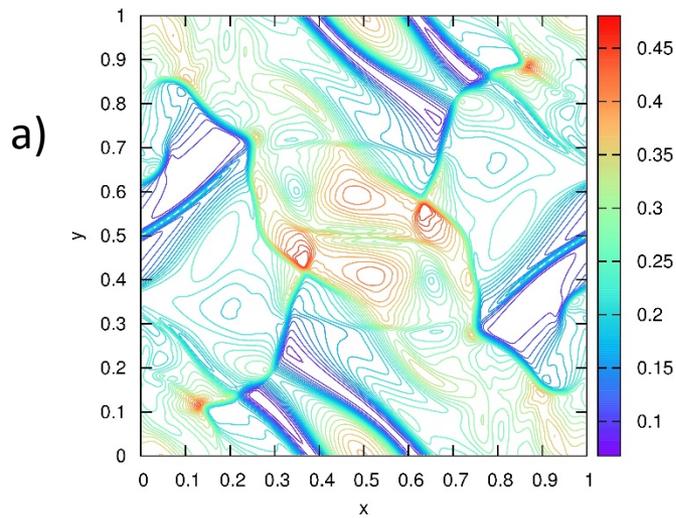
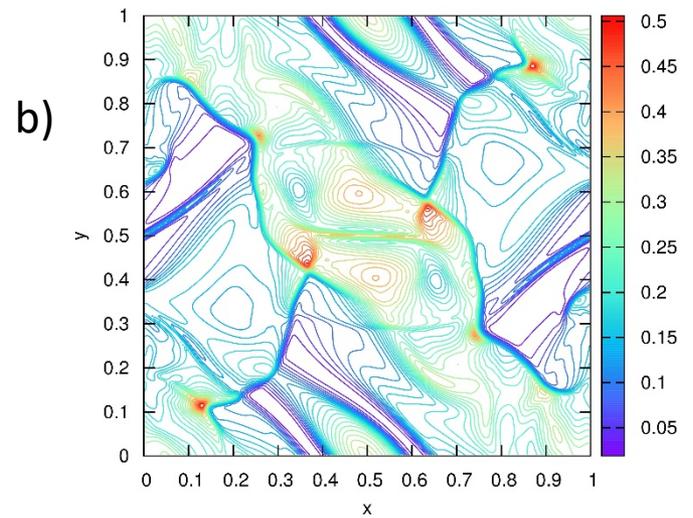
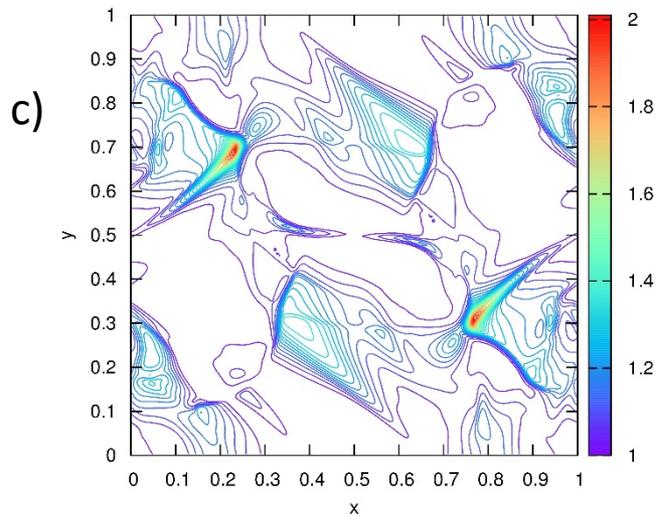
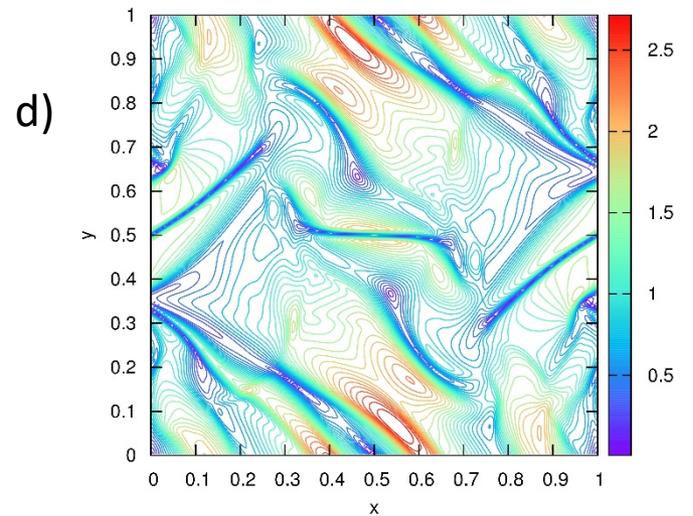

*Figs. 6a and 6b show the total density and total gas pressure. Figs. 6c and 6d show the Lorentz factor for the positrons (the electrons have a comparable Lorentz factor) and the magnitude of the magnetic field. The spatially fourth order accurate scheme was used.*

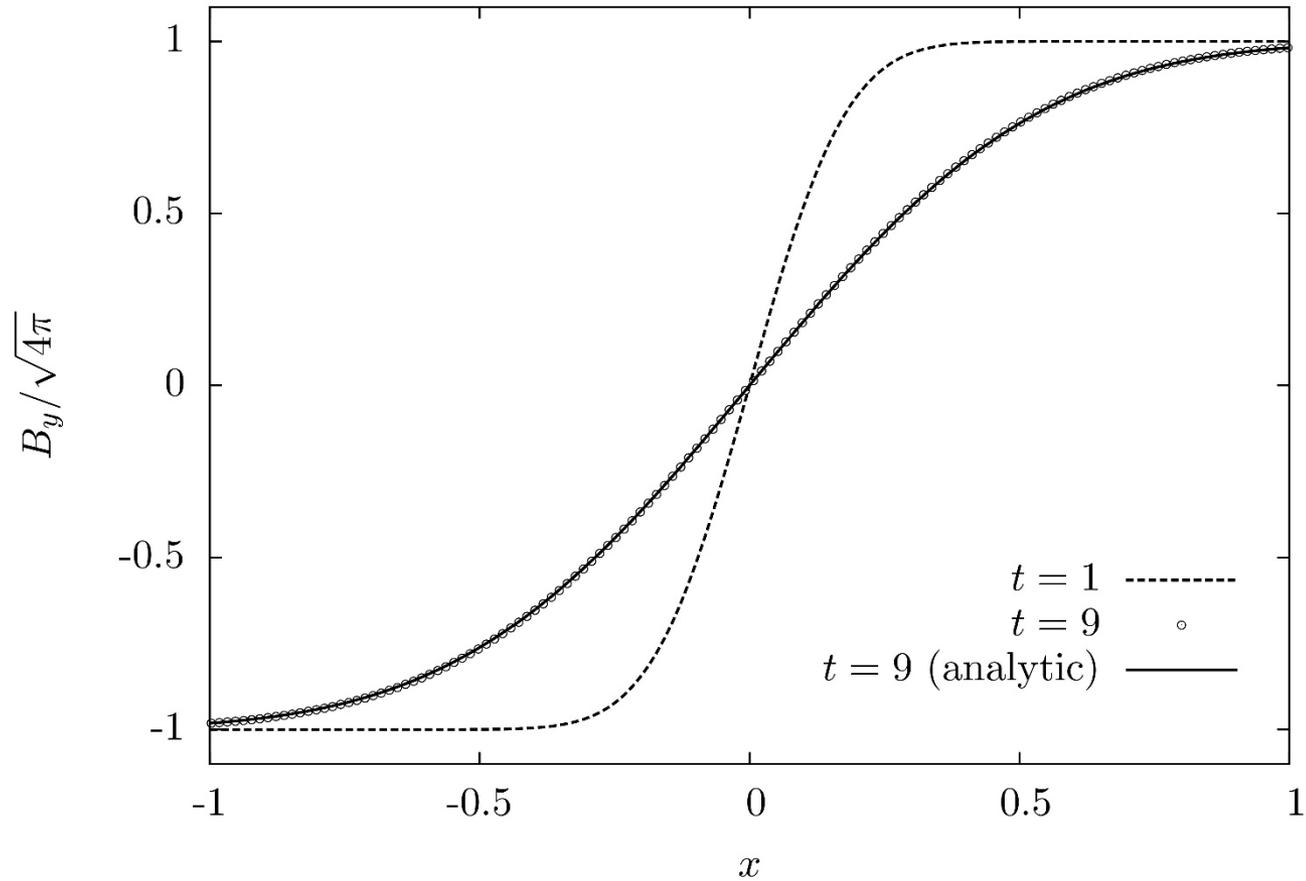

Fig 7 shows y-magnetic field at initial (dashed) and final time for the current sheet problem with finite resistivity. The numerical solution at the final time (dot) exactly matches with analytic solution at the final time (solid line).

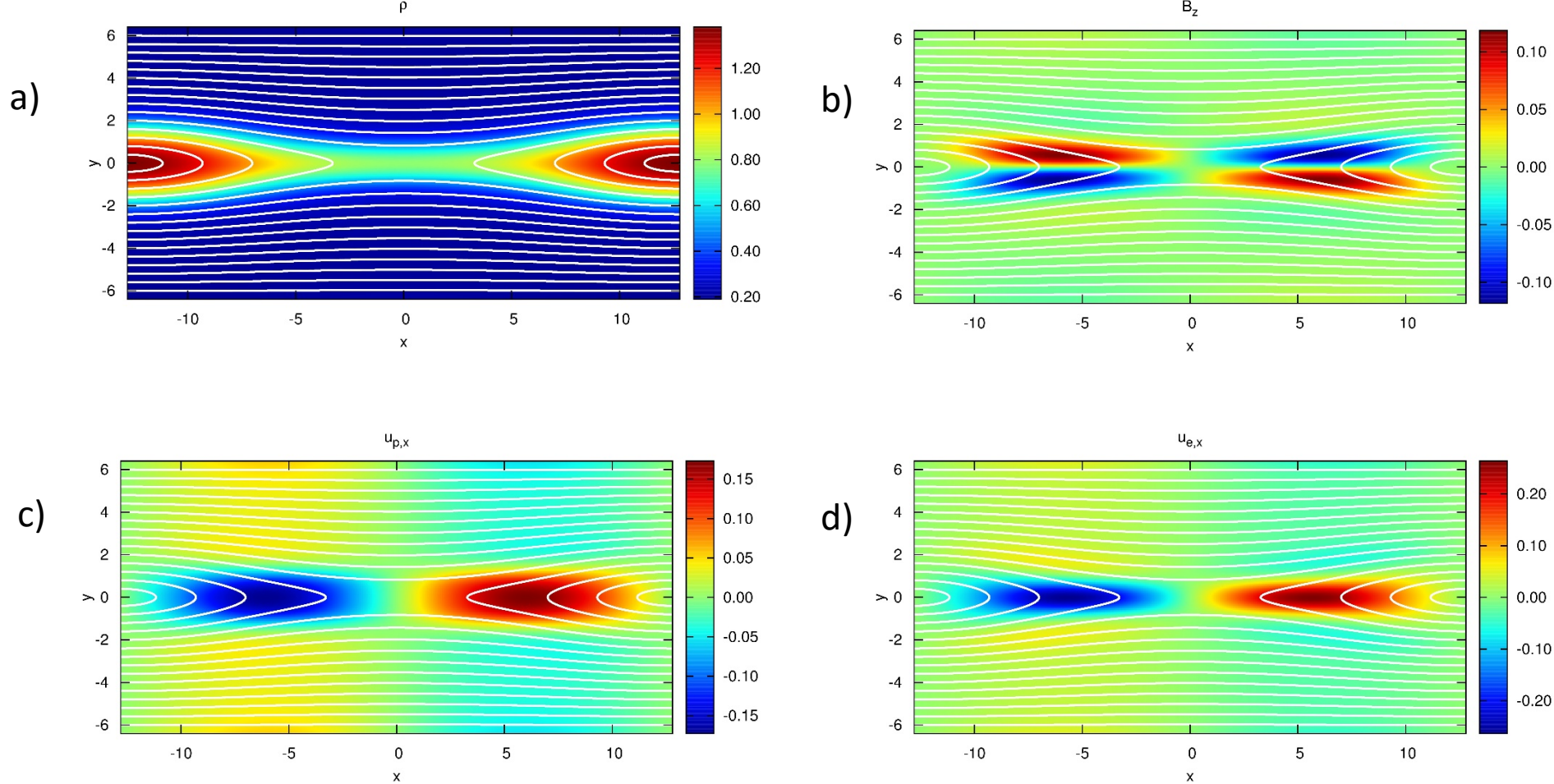

*Fig. 8a shows the total mass density, Fig. 8b shows the z-component of the magnetic field. Figs. 8c and 8d show the x-components of four-velocity for the protons and electrons respectively. White contours show the overlaid magnetic field lines. Fig. 8 corresponds to a time of 40 units.*

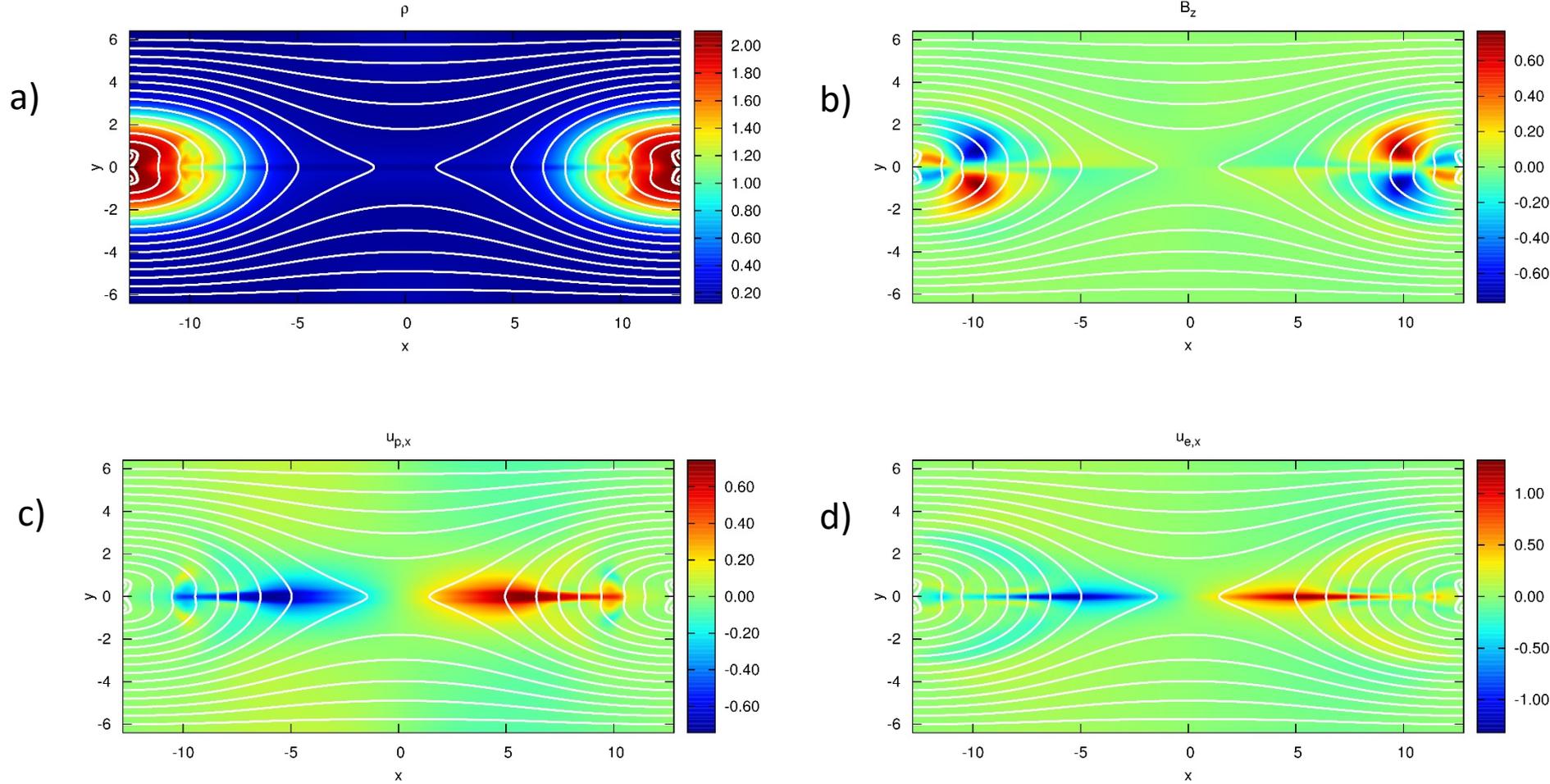

The panels in Fig. 9 show the same flow variables as Fig. 8, but at a time of 80 units.